\documentclass[%
 aip,
 amsmath,amssymb,
 reprint
]{revtex4-2}

\usepackage{graphicx}
\usepackage{dcolumn}
\usepackage{bm}

\usepackage[utf8]{inputenc}
\usepackage[T1]{fontenc}
\usepackage{mathptmx}
\usepackage{etoolbox}
\usepackage{amsmath}
\usepackage{xcolor}

\usepackage{hyperref}

\usepackage{color,soul}

\graphicspath{{graphs/}}
\def\scso/{Sr\textsubscript{2}CoSi\textsubscript{2}O\textsubscript{7}}
\def\cbco/{CaBaCo\textsubscript{4}O\textsubscript{7}}
\def\lcpo/{LiCoPO\textsubscript{4}}
\def\bcgo/{Ba\textsubscript{2}CoGe\textsubscript{2}O\textsubscript{7}}
\def\bfo/{BiFeO\textsubscript{3}}
\def\cnbo/{CoNb\textsubscript{2}O\textsubscript{6}}
\def\DM/{Dzyaloshinskii-Moriya}
\def\wn/{cm\textsuperscript{-1}}
\def\pd/{\emph{p}\,--\,\emph{d}}
\def\Tn/{\textit{T}\textsubscript{N}}
\def\Tc/{\textit{T}\textsubscript{C}}
\def\URS{URu$_{2}$Si$_{2}$\,}
\def\Bi2Se3{Bi$_2$Se$_3$\,}
\def\Sr2RuO4{Sr$_2$RuO$_4$\,}
\def\UPt3{UPt$_3$\,}
\def\POS{PrOs$_4$Sb$_{12}$\,}
\def\UTe2{UTe$_2$\,}
\def\edc/{\ensuremath{\mathbf{E}}}
\def\ddc/{\ensuremath{\mathbf{D}}}
\def\eac/{\ensuremath{\mathbf{E}_\omega}}
\def\dac/{\ensuremath{\mathbf{D}_\omega}}
\def\bdc/{\ensuremath{\mathbf{B}}}
\def\hdc/{\ensuremath{\mathbf{H}}}
\def\bac/{\ensuremath{\mathbf{B}_\omega}}
\def\hac/{\ensuremath{\mathbf{H}_\omega}}
\def\pac/{\ensuremath{\mathbf{P}_\omega}}
\def\mac/{\ensuremath{\mathbf{M}_\omega}}
\def\pdc/{\ensuremath{\mathbf{P}}}
\def\mdc/{\ensuremath{\mathbf{M}}}

\def\rotK/{\ensuremath{\Theta_\mathrm{K}}}

\newcommand{\vect}[1]{\ensuremath{\mathbf{#1}}}
\newcommand{\mat}[1]{\ensuremath{\hat{\mathbf{#1}}}}

\def\mob/{$\mathrm{cm}^2\mathrm{V}^{-1}\mathrm{s}^{-1}$}

\begin{document}

\title[]{Modified Martin-Puplett interferometer for magneto-optical Kerr effect measurements at sub-THz frequencies}

\author{A. Glezer Moshe}
 \email{aviv.glezer@kbfi.ee}
\affiliation{National Institute of Chemical Physics and Biophysics, Akadeemia tee 23, 12618 Tallinn, Estonia}
\author{R. Nagarajan}
\affiliation{National Institute of Chemical Physics and Biophysics, Akadeemia tee 23, 12618 Tallinn, Estonia}
\author{U. Nagel}
\affiliation{National Institute of Chemical Physics and Biophysics, Akadeemia tee 23, 12618 Tallinn, Estonia}
\author{T. R\~o\~om}
 \email{toomas.room@kbfi.ee}
\affiliation{National Institute of Chemical Physics and Biophysics, Akadeemia tee 23, 12618 Tallinn, Estonia}
\author{G. Blumberg}
 \email{girsh@physics.rutgers.edu}
\affiliation{National Institute of Chemical Physics and Biophysics, Akadeemia tee 23, 12618 Tallinn, Estonia}
\affiliation{Department of Physics and Astronomy, Rutgers University, Piscataway, New Jersey 08854, USA}

\date{\today}

\begin{abstract}
We present magneto-optical Kerr effect (MOKE) spectrometer based on a modified Martin-Puplett interferometer, utilizing  continuous wave sub-THz low-power radiation in broad frequency range. 
This spectrometer is capable of measuring the frequency dependence of the MOKE response function, both the Kerr rotation and ellipticity, simultaneously with sub-milliradian accuracy without the need for reference measurement. 
The instrument's versatility allow it to be coupled to a cryostat with optical windows enabling studies of a variety of quantum materials such as unconventional superconductors, two-dimensional electron gas systems, quantum magnets, and other system showing optical Hall response, at sub-Kelvin temperatures and in high magnetic fields. 
We demonstrate the functionality of the MOKE spectrometer using an undoped InSb wafer as a test sample. 

\end{abstract}

\maketitle
{}

\section{\label{Introduction}Introduction}


The application of a magnetic field can significantly alter the properties of materials. 
It is well-accepted that magneto-resistivity and Hall-effect measurements provide more comprehensive information compared to standard dc zero-field transport measurements. 
Similarly, the magneto-optical response carries new data about physical systems.\cite{Kimel2022}  

Recent advancements in magneto-optical techniques have allowed for the acquisition of novel information about electronic systems by extending dc Hall conductivity measurements into the THz frequency range.\cite{Levallois2015} 
This spectroscopic approach is designed to complement transport experiments, testing the bounds of standard relaxation time approximation in metals,\cite{Drew1997} measuring intrinsic Berry-phase contributions to the anomalous Hall effect, and probing mechanisms of spontaneous time-reversal symmetry breaking.\cite{Cho2016}  
Electronic systems where magneto-optical spectroscopy has recently proven to be particularly informative are unconventional superconductors,\cite{Kallin2017,Kapitulnik2018} the integer and fractional quantum Hall systems,\cite{Ikebe2010,Dziom2019} quantum anomalous Hall states,\cite{Okada2016} graphene-like systems,\cite{Gusynin2007} magnetic topological insulators,\cite{Fang2010,Xue2013} valley Hall effect 2D crystals,\cite{McEuen2014} the "strange metals" including cuprates and heavy fermions near antiferromagnetic quantum critical points,\cite{Drew1997} to name a few. 

Magnetic fields induce non-reciprocal optical phenomena, such as the Faraday and Kerr rotations. 
These magneto-optical rotations play a pivotal role in characterizing quantum materials. 
An examples of notable area of recent interest is chiral superconductors, which possess non-trivial topological properties and order parameters that spontaneously break time-reversal symmetry (TRS).\cite{Kallin2016,Sato2017} 

One of the greatest challenges in identifying superconductors that spontaneously break TRS is the limited availability of suitable probes.\cite{Ghosh2020}  
Optical measurements of the polar Kerr effect (PKE), where the rotation angle of linearly polarized radiation reflected from a sample’s surface is detected, have emerged as a key method for identifying TRS-broken states in unconventional materials.\cite{Kapitulnik2009}  
The Kerr rotation arises from the ac Hall conductivity in dimensions perpendicular to the light’s wave vector.  
For this Hall conductivity to be finite, the medium must break time-reversal symmetry and mirror symmetries about all planes parallel to the wave vector.\cite{Cho2016} 
The detection of non-vanishing PKE polarization rotation at zero external magnetic field provides unambiguous evidence for finite anomalous ac Hall conductivity, thus indicates TRS-breaking.\cite{Yakovenko2019PRX,Cho2016,Kapitulnik2015notes}  

To distinguish the signature of broken TRS from effects due to natural optical activity, it is critical to measure the Kerr rotation angle rather than the Faraday rotation. 
This distinction is necessary because PKE is absent if TRS is preserved.\cite{Cho2016} 

Although recent PKE studies using a telecom laser at approximately a micrometer wavelength have identified spontaneous Kerr responses in several unconventional superconductors, such as 
\UPt3,\cite{Kapitulnik2014} \URS,\cite{Kapitulnik2015u} \POS,\cite{Kapitulnik2018} \UTe2,\cite{Hayes2021} FeTe$_{0.55}$Se$_{0.45},$\cite{Farhang2023} 
a significant limitation of these experiments is their restriction to a single near-visible radiation frequency. 
This frequency far exceeds the energy scale relevant to unconventional superconductivity, typically around 0.1 THz. Consequently, these results have sparked considerable debates regarding their interpretation. 
Key questions include whether the observed effects are intrinsic bulk properties of the superconductor or influenced by external factors such as pinned vortices,\cite{UTe2_Rosa2023,UTe2_RosaPRL2023} or inhomogeneities of the sample. 
More importantly, the microscopic origin and quantitative understanding of the TRS-broken state remain contentious issues that can only be resolved through ultra-low-frequency spectroscopic measurements of the Kerr response as a function of frequency, within the range relevant to superconductivity and magnetism that causes TRS-breaking. 
This necessitates a new generation of spectroscopic THz-range magneto-optical Kerr effect (MOKE) instruments.

There are very few MOKE instruments operating in the THz range. 
R. Shimano \textit{et al.}\cite{Shimano2002,Ino2004} used a time-domain THz setup to measure the amplitude of reflected radiation, with an angle of incidence of 45 degrees in a 0.48 T magnetic field at room temperature, achieving an accuracy of about 10\,mrad between 0.5 and 2.5\,THz. 
The setup of D. Drew \textit{et al.}\cite{Jenkins2010rsi, Jenkins2010prb} employs a CO$_2$ laser to generate THz radiation at selected frequencies of 0.7, 1.3, and 2.5\,THz. 
Using a rapidly rotating waveplate, the Faraday rotation angle can be measured with a resolution of about 0.1\,mrad in transmission, and the Kerr rotation angle can be measured with a resolution of about 1\,mrad in reflection. 
In time-domain methods, it is possible to separate Kerr and Faraday effects by isolating the Faraday rotation from the first pulse and subtracting it from the reflected pulse that travels twice through the sample and arrives later in time. 
The most recent time-domain setup, which employs dual detectors, reports a Faraday angle precision of about 0.02\,mrad.\cite{Tagay2024}  
The challenges of THz-range ellipsometry are comprehensively discussed elsewhere.\cite{Neshat2013}

The previously described methods either use high-power radiation or are limited in the selection and resolution of frequencies. 
High-power radiation is highly detrimental for studying samples at sub-Kelvin temperatures, necessitating a new approach.

We have developed a MOKE spectrometer that uses very low incident power on the sample, about 0.1-1\,$\mu$W, in the range of 0.1-1\,THz to simultaneously measure both the real and imaginary parts of the Kerr response function with sub-milliradian accuracy. 
Below, we explain the underlying principle of the interferometer-based approach and demonstrate the performance of the instrument using MOKE measurement in an InSb semiconductor as a test sample.

\section{\label{sec:Kerr} Polar Kerr effect}

\subsection{Reflection matrix transformation}

Here we analyse the state of initially linearly polarized radiation after reflection under normal incidence from a material where the normal modes of electromagnetic radiation propagation are circularly polarized.
The relation between linearly polarized radiation,  $\vect{E} = (E_x, E_y)$, and circularly polarized radiation propagating along the $z$ axis  is 
\begin{equation}\label{eq:lin2circ}
  (E_-,E_+)=\mat{U} (E_x,E_y),  
\end{equation}
where the transformation matrix is 
\begin{eqnarray}\label{eq:U_iso}
	\mat{U} &=&\frac{1}{\sqrt{2}} 
	\left( 
	\begin{array}{cc}
		1 & -i\\
		-1 & -i
	\end{array}
	\right).
\end{eqnarray}

The reflection coefficient from the material-vacuum interface is 
\begin{equation}\label{eq:rpm_matrix}
	\mat{r}=\left(
	\begin{array}{cc}
		r_- & 0\\
		0 & r_+
	\end{array}
	\right).
\end{equation}
Using the complex indexes of refraction for the right-handed ($N_-$) and left-handed ($N_+$) modes in the material the reflection coefficients for the vacuum-material interface are
\begin{equation}\label{eq:rpm_normal_inc}
	r_\pm= \frac{N_\pm -1 }{N_\pm +1 }.
\end{equation}

The initially linearly polarized radiation $\vect{E}^i$ after reflection is transformed into $\vect{E}^f=(E_x^f,E_y^f)$, 
\begin{eqnarray}\label{eq:ein_eout}
	\vect{E}^f& =& \left(\begin{array}{cc}
	-1& 0\\
	0& 1
	\end{array}\right)
	 \mat{U}^{-1} \mat{r} \mat{U}  \vect{E}^i =\mat{r}_{xy}  \vect{E}^i,
\end{eqnarray}
where the matrix 
\begin{eqnarray*}\label{eq:11}
	& & \left(\begin{array}{cc}
	-1& 0\\
	0& 1
	\end{array}\right)
\end{eqnarray*}
inverts the $x$ axis of the coordinate frame.
Such transformation of the coordinate frame also changes the  handedness of the reflected radiation what is defined relative to the propagation direction:
\begin{equation}\label{eq:R2L}
	(E_+,E_-)= \mat{U}  \left(\begin{array}{cc}
		-1& 0\\
		0& 1
	\end{array}\right)  \mat{U}^{-1} (E_-,E_+).
\end{equation}
From Eq.\,\ref{eq:ein_eout} we get the reflection matrix for the linearly polarized radiation,
\begin{eqnarray}\label{eq:r_matrix_xy}
	\mat{r}_{xy} & =& -\frac{1}{2}\left(\begin{array}{cc}
		r_+ + r_-& i (r_+ -r_-)\\
		 i (r_+ -r_-)& -(r_+ + r_-)
	\end{array}\right).
\end{eqnarray}

\begin{figure}
	\centering
	\includegraphics[width=0.7\linewidth]{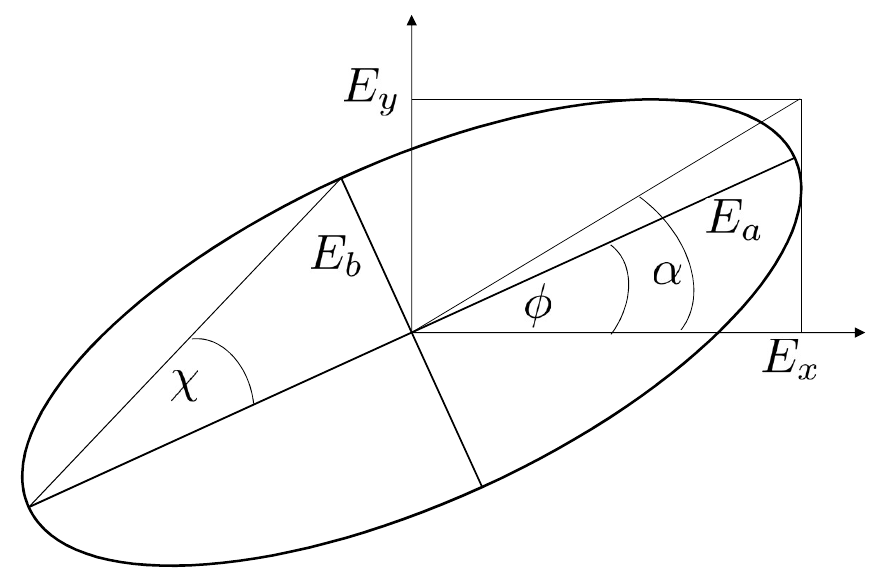}
	\caption{\label{fig:ellipse}
	The length of ellipse's  axes are $E_a$ and $E_b$ and the major axis is rotated from the $x$ axis by $\phi$.
	The electric field in the $xy$ coordinate frame is $(E_x,E_y e^{i\theta})$ where $E_x$ and $E_y$ are the amplitudes and $\theta$ is the phase difference between the components. 
	}
\end{figure}


\subsection{Kerr rotation and ellipticity}


Lets assume the incident  linearly polarized radiation is  $\vect{E}^i=(E_x^i,0)$.
Then, the reflected radiation using  Eq.\,\ref{eq:r_matrix_xy} becomes,
\begin{equation}\label{eq:EyEx}
	(E_x^f,E_y^f e^{i\theta})=
	-\frac{E_x^i}{2}\left(
	\begin{array}{c}
		r_+ + r_-\\
		i(r_+ - r_-),  
	\end{array}
	\right)
\end{equation}
where $\theta$ is the phase difference between $y$ and $x$ components of the reflected radiation.
We define the complex  Kerr response function $\rotK/$ as 
\begin{equation}\label{eq:rot_from_pm_r}
\rotK/ \equiv \Phi + i \mathrm{X}=\frac{E_y^f}{E_x^f}e^{i\theta} = \frac{i(r_+ - r_-)}{	r_+ + r_-},
\end{equation}
where the last equality comes from (\ref{eq:EyEx}).
$\Phi$ and $\mathrm{X}$ are the real and the imaginary parts of \rotK/.
With  (\ref{eq:rot_from_pm_r})  the reflection matrix (\ref{eq:r_matrix_xy}) for the linearly  polarized radiation reads:
\begin{eqnarray}\label{eq:r_matrix_Ex}
	\mat{r}_{xy} & =& -\frac{r_+ + r_-}{2}\left(\begin{array}{cc}
		1& \rotK/\\
		\rotK/ & -1
	\end{array}\right),
\end{eqnarray}
The prefactor $(r_+ +r_-)/2=-E_x^f/E_x^i$  is the  complex reflection  coefficient for the linearly polarized radiation.

The Kerr response function \rotK/ is related to the rotation and ellipticity of radiation as follows.
An elliptically polarized radiation, $ (E_x, E_y \exp(i \theta))$, is characterized by two parameters,
the phase difference $\theta$, and the ratio of two orthogonal amplitudes, $\tan \alpha  =  E_y/E_x$. 
With this notation
\begin{equation}\label{eq:elliptical_general}
	\rotK/ =\tan\alpha (\cos \theta + i \sin \theta).
\end{equation}
The  elliptically polarized state  can be transformed into a coordinate frame where the phase difference between two orthogonal components is $\pi/2$,  $\vect{E}= (E_a, E_b \exp(i \pi/2))$,   see Fig.\,\ref{fig:ellipse}.
The rotation of the ellipse main axis from the $x$ axis is $\phi$ and the ellipticity  is $	\chi =\arctan(E_b/E_a) $.
Then, see e.g.,\cite[p. 28]{BornWolf}
\begin{eqnarray}\label{eq:ellipse_relations}
	\tan(2\phi) &=& \tan(2\alpha)\cos\theta,\\
	\sin(2\chi) &=& \sin(2\alpha) \sin\theta.\nonumber
\end{eqnarray} 

If the ratio of amplitudes is small, $\alpha \ll 1$, we get from  (\ref{eq:elliptical_general}) and (\ref{eq:ellipse_relations}) the relation between the rotation  and the ellipticity of radiation and the real and imaginary parts of the Kerr response function: 
\begin{eqnarray}\label{eq:small_ellipse}
	\phi &=& \alpha \cos\theta = \Phi,\\
	\chi &=& \alpha \sin\theta = X.\nonumber
\end{eqnarray}

Under  the normal incidence reflection  from the vacuum-material interface,  Eq.\,(\ref{eq:rpm_normal_inc}),    the Kerr response function, Eq.\,\ref{eq:rot_from_pm_r}, depends on the  material index of refraction as:
\begin{equation}\label{eq:rot_from_N_pm}
	\rotK/ = \frac{i(N_+ - N_-)}{	N_+ N_- -1}.
\end{equation}
In the small $\alpha$ limit using (\ref{eq:rot_from_N_pm}) in (\ref{eq:small_ellipse}) we recover Argyres's result  for the Kerr rotation and ellipticity:\cite{Argyres1955} 
\begin{eqnarray}\label{eq:Argyres}
	\phi &=&  -\Im  \frac{(N_+ - N_-)}{	N_+ N_- -1},\\
	\chi &=&  \Re  \frac{(N_+ - N_-)}{	N_+ N_- -1}.\nonumber
\end{eqnarray}

\section{\label{sec:model}Working principle of the MOKE spectrometer }
\subsection{\label{sec:MP_interferometer}The modified Martin-Puplett interferometer}

\begin{figure*}[t]
\centering
	\includegraphics[width=1.8\columnwidth]{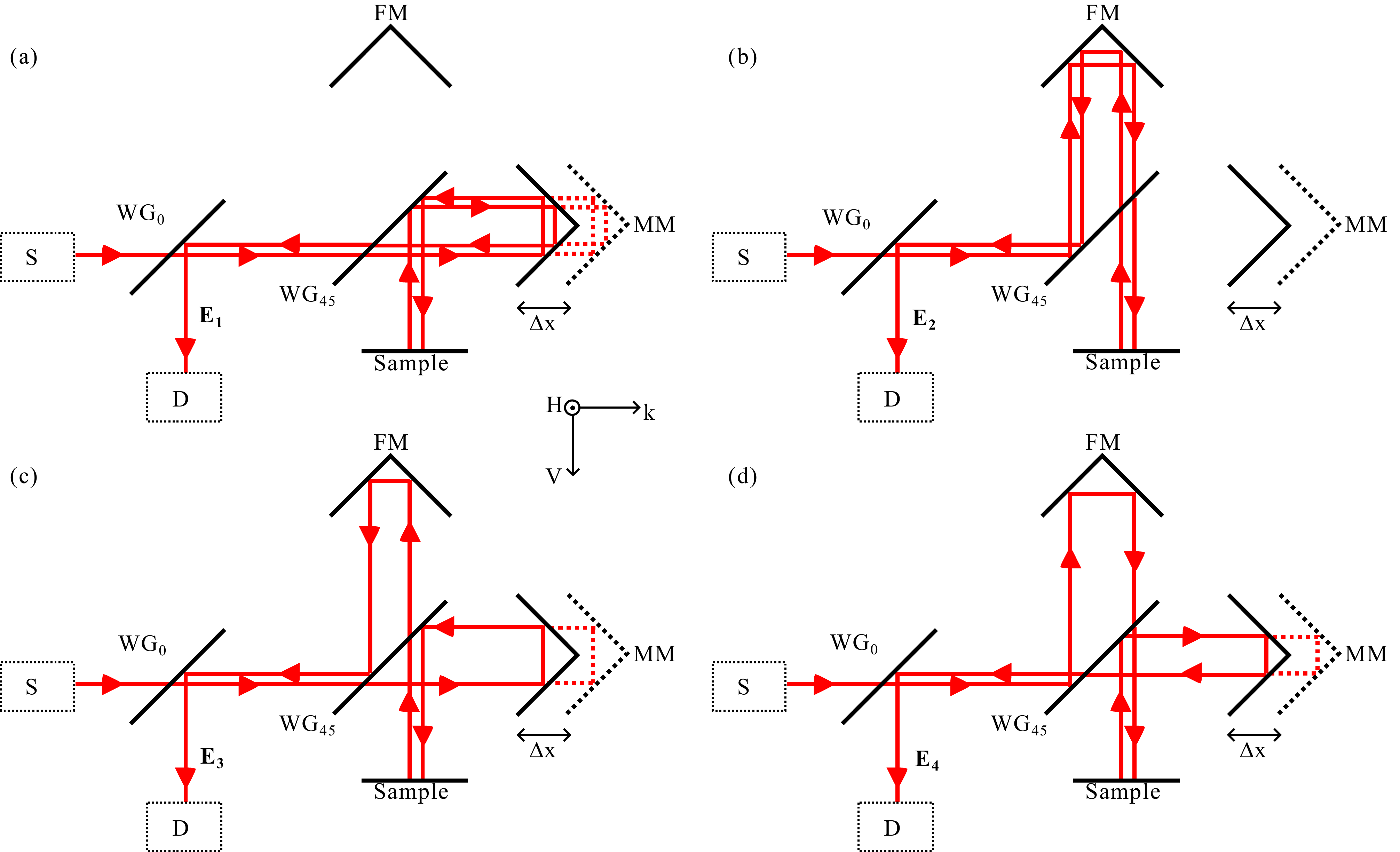}
	\caption{\label{fig:4Beams}
    Schematic of the MOKE interferometer and the radiation path through the MPI described in Eq.\,\ref{eq:4beams}. 
    The coordinate axis in the figure is in frame of reference of radiation beam emitted from source. 
    H and V refer to horizontal and vertical direction of polarization of radiation propagating along $\vect{k}$.
    The optical elements are source (S), detector (D), fixed roof top mirror (FM) and moving roof top mirror (MM). 
    WG$_{\eta}$ is the wiregrid polarizer with wires aligned at angle $\eta$  with respect to the H-axis shown in the figure. 
    The displacement of the MM is $\Delta x$, where $\Delta x=0$ is the MM position when it is at the same distance from the beam splitter WG$_{45}$ as FM. 
    The direction of wires of WG$_{45}$ is $\eta=45^\circ$ when the beam is incident on the beamsplitter from the left or from the top  in figure. 
    For the beams incident on the beamsplitter from the right or from the bottom $\eta=-45^\circ$. 
    The roof top mirror flips the polarization of the incident radiation as described by the Jones matrix in Table\,\ref{tab:Jones_matrix}. 
    If the incident radiation has polarization at $\pm 45^\circ$ relative to the roof line the reflected radiation will be orthogonal to the incident radiation. 
    (a) The radiation beam is reflected from MM to sample to MM, adding  the path difference $4 \Delta x$, as described by $E_1$ in Eq.\,\ref{eq:4beams}. 
    (b) The radiation beam is reflected from FM to sample to FM and path difference for this beam is zero, as described by $E_2$. 
    (c,d) The radiation beam is reflected from MM (FM) to sample to FM (MM) respectively, adding a path difference $2 \Delta x$, as described by $E_3$ ($E_4$). }
\end{figure*}

A schematic of the polar MOKE spectrometer is shown in Fig.\,\ref{fig:4Beams}.
The central part of the spectrometer is the modified Martin-Puplett interferometer (MPI) \cite{Martin1970} consisting of a wiregrid beamsplitter WG$_{45}$ and two roof top mirrors: the fixed mirror (FM) and the moving mirror (MM).
To enable the polar Kerr spectroscopy under normal incidence a sample is placed into one of the two ports of MPI, and the source and detector, separated by the wiregrid WG$_{0}$, are in the other port \footnote{Interferometer arms are the ones with roof mirrors with the beamsplitter in between. The modified MP interferometer has two symmetric ports on the other side of the beamsplitter: input and output.}.

The beam emitted from the source S after passing through the polarizer $\mathrm{WG_{0}}$ is vertically polarized, $\vect{E}_0=E_0(1,0)$.
The beamsplitter WG$_{45}$ splits the beam into two orthogonally polarized beams.
These two beams, $\vect{E}_1$ and $\vect{E}_2$, after reflection from the sample arrive to detector D as shown in Fig.\,\ref{fig:4Beams}\,(a) and (b).
If the sample has a non-zero MOKE the beam after the reflection from the sample is split again by the beamsplitter WG$_{45}$ and two additional beams, $\vect{E}_3$ and $\vect{E}_4$,  reach the detector D, Fig.\,\ref{fig:4Beams}\,(c) and (d).
The wiregrid WG$_{0}$ transmits the vertical polarization from the source into the interferometer and reflects the horizontal polarization from the interferometer output beam to the detector.

The four beams reaching the detector D are derived by applying the Jones matrices, Table\,\ref{tab:Jones_matrix}, to the source beam  $\vect{E}_0$:
\begin{widetext}
     \begin{eqnarray}\label{eq:4beams}
\vect{E}_1& = & \mathrm{WG}_{0}^\mathrm{R}\cdot \mathrm{WG}_{-45}^\mathrm{T}\cdot\mathrm{MM}\cdot\mathrm{WG}_{-45}^\mathrm{R}\cdot\mathrm{R}_\Theta\cdot\mathrm{WG}_{-45}^\mathrm{R}\cdot\mathrm{MM}\cdot\mathrm{WG}_{45}^\mathrm{T}\cdot \mathrm{WG}_{0}^\mathrm{T}\cdot E_0(0,1)= -\frac{1}{2} E_0 \bar{r} e^{8 i \pi \xi} (1,0) ,\nonumber\\ 
\vect{E}_2&=& \mathrm{WG}_{0}^\mathrm{R} \cdot \mathrm{WG}_{45}^\mathrm{R}\cdot\mathrm{FM}\cdot\mathrm{WG}_{-45}^\mathrm{T}\cdot\mathrm{R}_\Theta\cdot\mathrm{WG}_{45}^\mathrm{T}\cdot\mathrm{FM}\cdot\mathrm{WG}_{45}^\mathrm{R}\cdot \mathrm{WG}_{0}^\mathrm{T}\cdot E_0(0,1)=\frac{1}{2}E_0 \bar{r} (1,0) , \nonumber\\
\vect{E}_3 &=& \mathrm{WG}_{0}^\mathrm{R} \cdot \mathrm{WG}_{45}^\mathrm{R}\cdot\mathrm{FM}\cdot\mathrm{WG}_{-45}^\mathrm{T}\cdot\mathrm{R}_\Theta\cdot\mathrm{WG}_{-45}^\mathrm{R}\cdot\mathrm{MM}\cdot\mathrm{WG}_{45}^\mathrm{T}\cdot \mathrm{WG}_{0}^\mathrm{T}\cdot E_0(0,1)=-\frac{1}{2} E_0 \bar{r} e^{4 i \pi \xi}\Theta_\mathrm{K} (1,0), \nonumber\\ 
\vect{E}_4& = &\mathrm{WG}_{0}^\mathrm{R} \cdot \mathrm{WG}_{-45}^\mathrm{T}\cdot\mathrm{MM}\cdot\mathrm{WG}_{-45}^\mathrm{R}\cdot\mathrm{R}_\Theta\cdot\mathrm{WG}_{45}^\mathrm{T}\cdot\mathrm{FM}\cdot\mathrm{WG}_{45}^\mathrm{R}\cdot \mathrm{WG}_{0}^\mathrm{T}\cdot E_0(0,1)=-\frac{1}{2} E_0 \bar{r} e^{4 i \pi \xi}\Theta_\mathrm{K} (1,0),
 \end{eqnarray}
\end{widetext}
here $\xi = \Delta x/\lambda$ is the reduced path difference where $\Delta x$ is the  position of the moving mirror relative to zero path difference (ZPD) position, $\lambda$ is the wavelength, and $\bar{r} = (r_+ + r_-)/2$.
After adding up the four beams in Eq.\,\ref{eq:4beams}   the  electric field on the detector is: 
 \begin{equation}\label{eq:Kerr_func}
    E(\xi)= E_0\bar{r} \left[ \frac{1}{2}\left( 1 - e^{8 i \pi \xi} \right) - e^{4 i \pi \xi} \rotK/ \right].
 \end{equation}

\begin{table}[b]
     \caption{
     \label{tab:Jones_matrix}
     Jones matrices of wiregrid, roof top mirror and back reflection from the sample exhibiting polar Kerr effect.
     $\eta$ is the angle of wiregrid wires and roof top mirror roof line from the horizontal direction measured towards the vertical direction.
     Horizontally polarized field has unit vector   $\vect{e}_h=(1,0)$ and vertically polarized has $\vect{e}_v=(0,1)$ and the beam propagation direction is  $\vect{e}_k=\vect{e}_h\times \vect{e}_v$.
     Sample reflection matrix is Eq.\,\ref{eq:r_matrix_Ex} where  $\bar{r} = (r_+ + r_-)/2$ and the complex  Kerr response function $\rotK/=\Phi + i \mathrm{X}$. 
}
\vspace{2mm}
		\begin{tabular}{|c|c|c|}
			\hline
			\textbf{Optical element}&\textbf{Jones matrix}&\textbf{ Symbol}\\
			\hline
			WG reflection & $\begin{pmatrix}
			-\cos^2\eta  & -\cos \eta \sin \eta \\
			\cos \eta \sin \eta & \sin^2\eta
		\end{pmatrix}$ & $\mathrm{WG}^R_\eta$\\
			WG  transmission & $	\begin{pmatrix}
			\sin^2\eta & -\cos \eta \sin \eta \\
			-\cos \eta \sin \eta & \cos^2\eta
		\end{pmatrix}$ & $\mathrm{WG}^T_\eta$\\
			Roof top mirror & $
			\begin{pmatrix}
			-\cos 2\eta &  -\sin 2\eta \\
			\sin 2\eta & -\cos 2\eta
		\end{pmatrix}$ & $\mathrm{RM}_\eta$\\
			Sample reflection & $-\bar{r} 
   \begin{pmatrix}
   1 & \rotK/ \\
   \rotK/ & -1 
   \end{pmatrix}$ & $\mathrm{R}_\Theta$\\
			\hline
		\end{tabular}
\end{table}

\begin{figure}[t]
	\centering
	\includegraphics[width=0.9\linewidth]{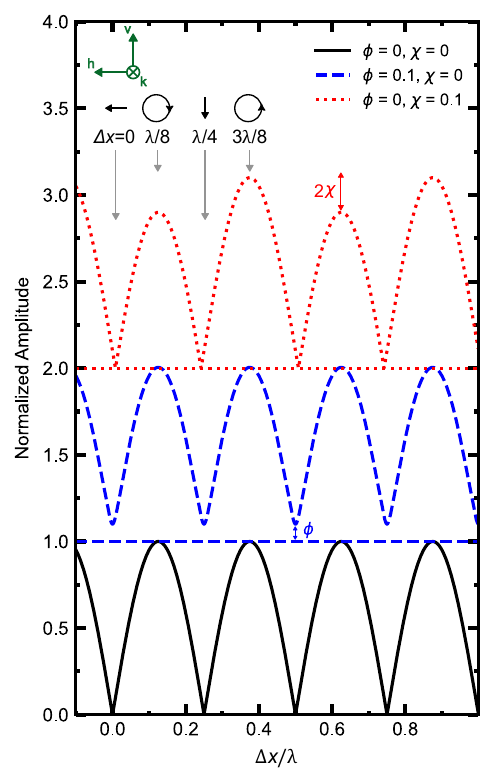}
	\caption{\label{fig:Int_Ideal_plots}
	Theoretical normalized electric field  amplitude $|E/E_0\bar{r}|$ at the detector calculated with Eq.\,\ref{eq:Kerr_func} for different values of \rotK/, arranged from bottom to top: zero rotation and zero ellipticity, finite rotation and zero ellipticity, and zero rotation and finite ellipticity, respectively.  
	The values in the legend are given in units of radians. 
	The curves are shifted by unity for clarity.
It can be seen that the rotation lifts the minima by $\phi$, while the ellipticity separates the maxima by $2\chi$. 
Special points in the interferogram correspond to the polarization state of the incident radiation on the sample. 
At the maxima, the radiation is either left or right circularly polarized, while at the minima, the radiation is linearly polarized along the horizontal or vertical direction. 
The direction of rotation and polarization are shown along with the local coordinate system.}
\end{figure}

To demonstrate how the real and imaginary parts of \rotK/ affect the interferogram we plot using Eq.\,\ref{eq:Kerr_func} the normalized amplitude $|E(\xi)/(E_0 \bar{r})|$ as a function of mirror displacement in Fig.\,\ref{fig:Int_Ideal_plots}.
At $\xi = 0$, the incident radiation at the sample surface is linearly polarized parallel to the direction transmitted by WG$_{0}$ and radiation reaches the detector only if the plane of polarization is rotated upon back reflection from the sample ($\phi \ne 0$): the real part of \rotK/ lifts the minima of the normalized amplitude.
At $\xi = 1/8$ or $\xi = 3/8$ we have right- or left-handed circularly polarized radiation incident on the sample surface and if an ellipticity $\chi \ne 0$ is introduced by the sample, the maxima in the detected interferogram have different amplitudes.
Finite ellipticity also shifts the positions of minima away from the position in the absence of ellipticity, $\xi= n /4$, where  $n=0, 1, 2, \ldots$. 

\begin{figure}[t]
	\centering
	\includegraphics[width=0.9\linewidth]{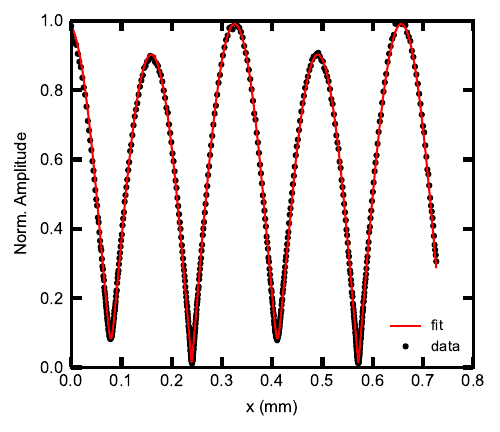}
	\caption{\label{fig:Interferogram_InSb}
     An example of an interferogram used to determine the rotation $\phi$ and ellipticity $\chi$ measured from undoped InSb semiconductor at $T=50$\,K and magnetic field 5\,mT.
    From the fit to Eq.\,\ref{eq:deltaR}, we obtained the rotation, $\phi = 36 \pm 1 $\,mrad, ellipticity, $\chi = 46.8 \pm 0.4$\,mrad, and the frequency $452.45\,\pm\,0.07$\,GHz.
    The radiation electric field amplitude, the data points, were fitted to a model (Eq.\,\ref{eq:deltaR}) that accounts for imbalance between interferometer arms $\delta = -0.0262 \pm 0.0006$.
    To measure the locations of the minima, the path difference step size $\Delta x$ was decreased linearly with decreasing amplitude down to 0.10\,$\mu$m from the initial value 3.37\,$\mu$m.
    The 1\,s integration time at maxima was increased as the inverse square root of the step size up to 5.7\,s per interferogram point.
    This strategy allowed us to limit the total measurement time of the interferogram with 612 points to 33\,minutes (including time for moving the roof mirror) while still having a relative amplitude error for each data point equal to about 0.5\,\%.}
\end{figure}

\subsection{\label{sec:data_analysis}Data analysis and accuracy}

Following Eq.\,\ref{eq:Kerr_func}, we show that the normalized electric field intensity for the interferogram acquires simple form which is the sum of unmodulated rotation induced intensity $\Phi^2$ and ellipticity induced baseline shift of interference modulation in the second intensity term: 
\begin{equation}\label{eq:Kerr_intensity}
    I(\xi) = 
    \left[\frac{E(\xi)}{E_0\bar{r}}\right]^2 = 
    \Phi^2 + [\mathrm{X} + \sin{(4 \pi \xi)}]^2.
  \end{equation}
  A difference of intensity interferograms shifted by a quarter of the reduced path difference enables direct measurement of ellipticity $\mathrm{X}$ from the normalized intensity modulation amplitude: 
 \begin{equation}\label{eq:Kerr_X_modulation}
    I(\xi) - I(\xi - \frac{1}{4}) = 4 \mathrm{X} \sin{(4 \pi \xi)}. 
   \end{equation}
Thus, one can obtain both the real and imaginary parts of the complex Kerr response function $\rotK/=\Phi + i \mathrm{X}$ from the fit to measured interferogram amplitudes, Eq.\,(\ref{eq:Kerr_func}), or intensities, Eqs.\,(\ref{eq:Kerr_intensity}-\ref{eq:Kerr_X_modulation}). 

The figure of merit for MOKE measurements is the accuracy of Kerr rotation angle $\delta\Phi$ and ellipticity $\delta\mathrm{X}$. 
Below we first analyze the theoretical error propagation for the fitting procedure to Eqs.\,(\ref{eq:Kerr_func}-\ref{eq:Kerr_intensity}) assuming normal distribution function for the radiation amplitude and the path difference position noise.\footnote{The latter can be consider as the interferogram phase fluctuation noise.}

We show below in Sec.\,\ref{sec:KerrToptica} that the implemented 
data acquisition procedure readily achieves measured relative signal amplitude fluctuations 
$\delta E_0/E_0$
better that $5 \times 10^{-3}$.\footnote{This accuracy can be further improved by implementing an amplitude modulation measurement schema.}
For the wavelength of interest, about a millimeter, commercially available high precision linear translation stages enable better than $10^{-4}$ accuracy for the relative path difference control $\delta \xi/\xi$. 
With these parameters, a relatively quick interferogram measurement containing $n = 200$ points per wavelength allows to obtain relative ellipticity accuracy 
$\frac{\delta\mathrm{X}}{\mathrm{X}} \propto \frac{\delta E_0}{E_0 \sqrt{n}} \approx 0.35 \times 10^{-4}$. 
Thus, small ellipticity can be quickly measured with sub-millirad accuracy, which can be further improved with enhancement of measurement points density.

The situation is less favorable for the Kerr rotation angle accuracy $\delta\Phi$ as this variable is not modulated by the interferogram scan, see Eq.\,(\ref{eq:Kerr_intensity}). 
It appears that the relative Kerr rotation angle accuracy is inversely proportional to the square root of number of points measured in the proximity of interferogram minima $n_{min}$, 
$\frac{\delta\Phi}{\Phi} \propto \frac{\delta E_0}{E_0 \sqrt{n_{min}}}$. 
Therefore, to obtain the Kerr rotation accuracy at the same order as the ellipticity accuracy, measurement with about an order of magnitude higher point density near the interferogram minima has to be implemented, see Sec.\,\ref{sec:KerrToptica}. 
Experimentally, the point density is limited by the minimal repeatable incremental move distance of translational stages.\footnote{Stages with air bearing and interferometric position readout provide better positioning accuracy.} 

Other contributions to the Kerr response function accuracy include photocurrent fluctuations in the Toptica detection system, interferences with intruding spurious back reflection beams from multiple surfaces of interferometer optical components, imbalance between two interferometer arms, frequency stability of the radiation, non-orthogonality of the sample surface to incoming beam, to name a few. 

For example, the effect of imbalance between two interferometer arms, which remains present even after careful alignment of the optical components, results in a deviation from the ideal interferogram shape shown in Fig.\,\ref{fig:Int_Ideal_plots}, making the odd and even interferogram minima unequal, see Fig.\,\ref{fig:Interferogram_InSb}. 
In practice, such imbalance between two interferometer arms can be described by a modified interferometer electric field amplitude function 
\begin{eqnarray}
   \label{eq:deltaR}
    E(\xi) &=& E_0\bar{r} \left[ 
    \frac{1}{2}
    \left( 
    \left(1 -\delta \right)^2 
    - e^{8 i \pi \xi} 
    \left(1 +\delta \right)^2 
    \right) - \right. \nonumber \\
& &    \left.-\, e^{4 i \pi \xi} \rotK/ 
    \left(1 -\delta^2 \right) \right], 
 \end{eqnarray}
where the electric field amplitude in the moving mirror arm is changed from the average of two by the factor $1+\delta$ and that of the fixed mirror arm by $1- \delta$. 
The Eq.\,\ref{eq:deltaR} introduces an additional, typically small, fitting parameter $\delta$, see Fig.\,\ref{fig:Interferogram_InSb}.

The relative signs of $\delta$ and $\Phi$ determine which of the two interferogram minima is smaller in amplitude.
If the signs are opposite then the even minima, including the one at ZPD, are deeper and with the same signs of  $\delta$ and $\Phi$ odd minima are deeper.
Thus, knowing the ZPD position and following the signs of $\delta$ and $\Phi$ across the spectrum we may obtain information about the sign of $\Phi$.\footnote{Additionally, the sign of $\Phi$ could be obtained from the behaviour of the phase of the detected signal. We have not used the phase information so far, because of too large overall phase drifts that are caused by THz frequency and optical setup instabilities, including the lasers and fibers that connect the lasers to the photomixers.}

We also note that $\rotK/(\omega)=\Phi(\omega) + i \mathrm{X}(\omega)$ is a spectroscopic response functions for which the real and imaginary parts are related by the Kramers–Kronig relations. 
Here $\omega$ is the radiation frequency. 
Thus, spectroscopic measurement of frequency dependence of ellipticity, $\mathrm{X}(\omega)$, which is faster than the direct Kerr angle $\Phi(\omega)$ measurement, may also give an estimate for the Kerr angle frequency dependence.

\section{\label{sec:results}Spectrometer and performance test}

\subsection{\label{sec:KerrToptica} Details of the MOKE spectrometer }

The MOKE spectrometer is comprised of three units: the radiation source, the modified MP interferometer, and the detector of interfered four beams. 

\begin{figure}[t]
	\centering
	\includegraphics[width=0.9\linewidth]{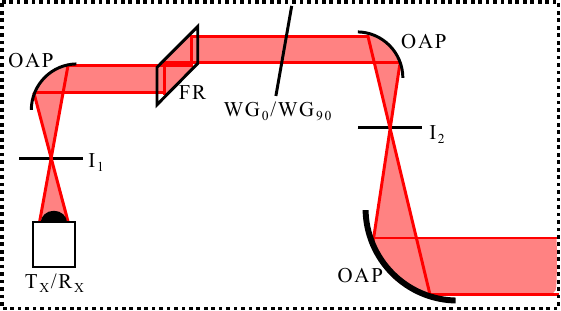}
	\caption{\label{fig:rhomb}
	Coupling of the transmitter/receiver ($\mathrm{T_X/R_X}$) to the modified MPI. 
	OAP and I are off-axis parabolic mirrors and iris apertures of the spacial filters to collimate the beam. 
	The Fresnel rhombs (FR) together with the output/input linear polarizers ($\mathrm{WG_{0}/WG_{90}}$) act as the broad band optical isolators of the source/detector units.}
\end{figure}

The optics assemblies for source and detector units are similar to each other, these are designed to 
({\em i}) collimate the incoming beam from transmitter $\mathrm{T_X}$ and focus the outgoing beam on the receiver $\mathrm{R_X}$, and 
({\em ii}) isolate spurious back reflections from re-entering into the interferometer, Fig.~\ref{fig:rhomb}.

The Toptica Terascan 780 frequency domain THz spectroscopy system with GaAs photomixers coupled to log-spiral antennas is used to generate and detect the continuous-wave (cw) THz radiation.
A voltage of about 10\,V is applied in the $\mathrm{T_X}$ to accelerate the charge carriers to generate tunable circularly polarized 0.1 -- 1\,THz radiation at the beat frequency of the two distributed feedback lasers.\cite{TopticaCW,Roggenbuck2010,Roggenbuck2013} 
We detect the photocurrent from $\mathrm{R_X}$ that is resulting from accelerating the charge carriers by the THz radiation.

An additional phase modulation extension, using two fiber stretchers, modulates the relative phase of laser radiation reaching the $\mathrm{T_X}$ and $\mathrm{R_X}$, enabling simultaneous detection of the amplitude and phase of the THz radiation.\cite{Roggenbuck2012stretcher} 
During a typical interferogram measurement time, about 30 minutes for 600 data points, the amplitude drifts less than 0.5\,$\%$.
The THz radiation frequency may drift over the interferogram measurement time: at 300\,GHz a typical frequency error from the fits is less than 0.08\,GHz.
The signal to noise ratio of detecting THz radiation amplitude depends on the integration time of the lockin detection at fiber stretcher frequency, where we typically reach more than 2 orders of magnitude in detecting the THz electric field amplitude.
Comparable signal to noise ratio is not achievable with bolometric detection, where intensity (amplitude squared) is being measured.

Time-domain THz systems also measure the THz radiation amplitude but they have a high THz radiation power on the sample surface. 
The advantage of time-domain system as compared to cw THz system is that it can separate the Fabry-Perot reflections  from the main signal, although at the cost of reduced spectral resolution.

To suppress back reflections into the interferometer we use broad band optical isolators in the source/detector units, comprising a linear polarizer and a Fresnel rhomb, see Fig.\,\ref{fig:rhomb}.
The Fresnel rhombs convert the polarization of radiation from circular/linear to linear/circular at the output/input of $\mathrm{T_X/R_X}$ so that the back reflected radiation from $\mathrm{T_X/R_X}$ has orthogonal polarization with respect to the original beam and therefore is reflected away from the interferometer by the linear polarizer $\mathrm{WG_{0}/WG_{90}}$.
We designed and made the rhombs from TOPAS cyclic olefin copolymer \cite{TOPAS} with non-parallel facets to avoid Fabry-Perot modulation within the rhombs.
With this we suppress the back reflections from $\mathrm{T_X/R_X}$ by more than an order of magnitude to less than 1\,\% level. 
The parallel THz beam incident on the Fresnel rhomb facet has 1" diameter, while the beam is expanded to 2" diameter for the modified MP interferometer.

We use wiregrid (WG) polarization beam-splitters from Specac which are made of $5~\mu$m diameter tungsten wires with the periodicity $12.5~\mu$m and have close to ideal transmittance and reflectance characteristics.\cite{Micica2016} 
At 0.3\,THz we measured better than 1:100 contrast in radiation amplitude between parallel and perpendicular polarized beams. 

The hollow rooftop mirrors comprise of two perpendicular (to 5 arc seconds accuracy) flat unprotected Au-coated mirrors,\cite{Rooftop} with the roof line aligned parallel to the H axis of the incident radiation beam, Fig.\,\ref{fig:4Beams}.
The bare gold coating of 150\,nm is sufficient for the frequency range of interest.

A 12\,mm long translation stage \cite{Thorlabs12mm} with calculated 29\,nm resolution of the actuator and 
minimal repeatable incremental movement 0.2\,$\mu$m 
was used to move the rooftop mirror to control the path difference in the modified MPI.

\subsection{\label{sec:performance} Performance test }

As it was first shown by Feil and Haas,\cite{Feil1987} the dispersion of the diagonal part of the dielectric constant near the plasma frequency resonance in a metallic system has a strong influence on the MOKE properties. 
The plasma edge leads to resonance-like peaks in the Kerr rotation and ellipticity spectra with strong enhancement of the MOKE magnitude. 
This prediction was verified for InAs by Shimano \textit{et al.}\cite{Shimano2002} at room temperature using a 0.48\,T permanent magnet. 

To test the performance of the MOKE spectrometer we carried out MOKE measurements from small direct gap undoped InSb semiconductor which upon cooling to 50\,K has low plasma frequency, 0.5\,THz, and high mobility, $\mu=3\times10^5$\,\mob/.  
The small effective mass of charge carriers in InSb, $m^*=0.014 m_e$, guarantees high cyclotron frequency $\omega_c=eB/m^*$ even for relatively low magnetic field $B$. 
Thus, the necessary condition of electron orbit coherence $\omega_c\tau > 1$ ($\tau$ is the mean free collision time) can be reached easily even in a few mT field. 
For this system Kerr rotation and ellipticity near the plasma frequency reaches 100\,mrad already at 5\,mT field.

\begin{figure}
	\centering
	\includegraphics[width=0.8\linewidth]{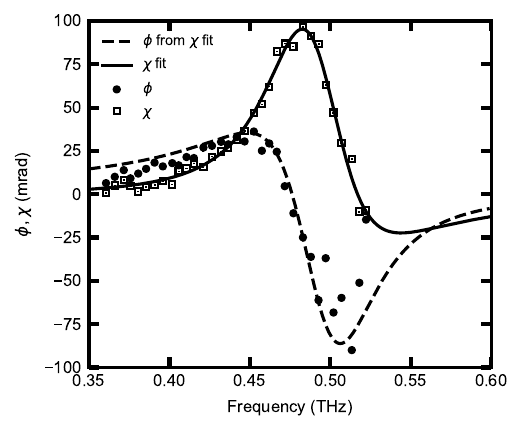}
	\caption{\label{fig:phi_chi_InSb}
    Rotation $\phi(\omega)$ and ellipticity $\chi(\omega)$ spectra measured from undoped InSb  at 50\,K and in magnetic field of 5\,mT. 
    To demonstrate the reliability of the measurement, we fitted the ellipticity $\chi$ (full squares) to the Drude model (solid line) and plotted the theoretical rotation $\phi$ (dashed line) using the fitting parameters obtained from the ellipticity fit. 
    The calculated rotation aligns well with the experimental data (full circles). The measurement time for each datapoint was 30 minutes.}
  \end{figure}

\begin{figure}
	\centering
	\includegraphics[width=0.8\linewidth]{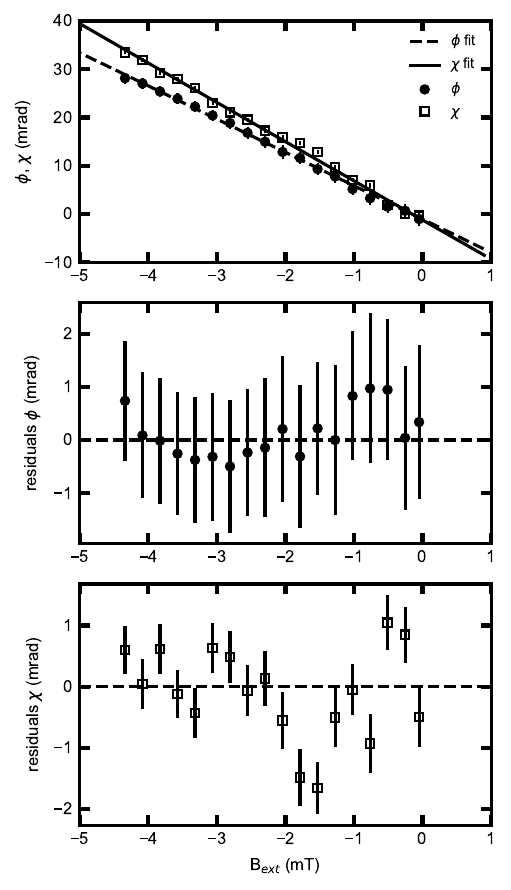}
 	\caption{\label{fig:phi_chi_InSb_Hdep}
    Magnetic field dependence of rotation $\phi$ and ellipticity $\chi$ at 442.2\,GHz and 50\,K in undoped InSb. 
    The dashed and solid lines are obtained by fitting simultaneously the magnetic field dependence of the rotation and ellipticity to the Drude model, Eq.\,\ref{eq:Drude}. The measurement time for each datapoint was 30 minutes.}
\end{figure}

Under external magnetic field perpendicular to the sample surface, $\vect{B}\parallel \vect{z}$,  the  optical conductivity tensor is \cite{Argyres1955}
\begin{eqnarray}\label{eq:r_matrix_sigma}
	\mat{\sigma} & =&\left(\begin{array}{cc}
		\sigma_{xx}& \sigma_{xy}\\
		-\sigma_{xy} & \sigma_{xx}
	\end{array}\right).
\end{eqnarray}
Within Drude approximation, the components of $\mat{\sigma}$ are:
\begin{subequations}\label{eq:Drude}
    \begin{align}
          \sigma_{xx} = \epsilon_0\epsilon_b\omega_p^2\frac{\gamma-i\omega}{(\gamma-i\omega)^2+\omega_c^2},\\
              \sigma_{xy} = -\epsilon_0\epsilon_b\omega_p^2\frac{\omega_c}{(\gamma-i\omega)^2+\omega_c^2},
    \end{align}
\end{subequations}
where $\omega_p=\sqrt{ne^2/m^*\epsilon_0\epsilon_b}$ is the plasma frequency, $n$ the carrier density, $e$ the electron charge, $\epsilon_0=8.85\times10^{-12}$\,F\,m$^{-1}$ the vacuum permittivity, $\epsilon_b = 15.68$ the background dielectric constant for InSb \cite{Howells1996}, and $\gamma = 1/\tau$ the scattering rate. In this case, it is convenient to work in a  circularly polarized radiation basis Eq.\,\ref{eq:lin2circ}, where the index of refraction $N$ depends on the handedness of radiation, 
\begin{subequations}\label{eq:eps_Npm}
    \begin{align}
        N_{\mp}=\sqrt{\epsilon_{xx} \pm i\epsilon_{xy}}, \\
        \epsilon_{xx}=\epsilon_{b} + i\frac{\sigma_{xx}}{\omega\epsilon_0}, \\
        \epsilon_{xy}= i\frac{\sigma_{xy}}{\omega\epsilon_0},
    \end{align}   
\end{subequations}
where we assumed $\mu=1$ for the magnetic permeability.
Using Eqs.\,\ref{eq:rot_from_N_pm}, \ref{eq:elliptical_general}, \ref{eq:ellipse_relations} and \ref{eq:eps_Npm}, we obtain the frequency-dependent rotation $\phi(\omega)$ and ellipticity  $\chi(\omega)$.
The three fitting parameters for the Drude model, Eq.\,\ref{eq:Drude}, are $\omega_p$, $\omega_c$ and $\gamma$.

As it is shown in Fig.\,\ref{fig:phi_chi_InSb}, the frequency dependencies of the Kerr angle rotation and ellipticity obtained by fits of the individual interferograms measured from undoped InSb wafer at 50\,K and the external magnetic field of 5\,mT using the modified interferometer electric field amplitude function, Eq.\,\ref{eq:deltaR},  are in agreement with the Drude model prediction. 
The result demonstrates the instrument's ability to track the sign change.
The ideal instrument, described by Eq.\,\ref{eq:Kerr_func} would not yield any information about the sign of $\phi$, but a small unbalance between the interferometer arms, Eq.\,\ref{eq:deltaR}, gives some information about the sign.
It can be seen that by fitting the ellipticity alone, the rotation is predicted quite well, thus proving the reliability of the measurement. 
The parameters obtained from the Drude fit are $\tau = 2.38\pm0.07$\,ps, $\omega_p/2\pi = 483.2 \pm 0.4$\,GHz, and $\omega_c/2\pi = 10.3 \pm 0.3$\,GHz. 
Using the relation $\mu=e\tau/m^*$, the mobility of the sample is $\mu = (2.99\pm0.07)\times10^{5}$ \mob/. 
These values are inline with time-domain THz spectroscopy (TDS) reflection measurements on undoped InSb,\cite{Howells1996} and with the dc transport measurements performed on a separate sample cut from the same InSb wafer.

In Fig.\,\ref{fig:phi_chi_InSb_Hdep} we also show the rotation and ellipticity as a function of magnetic field measured at $T=50$\,K and at frequency  $\omega=442$\,GHz. 
The magnetic field dependence of the rotation and ellipticity was simultaneously fitted to Drude model, Eq.\,\ref{eq:Drude}. 
The obtained fitting parameters, $\tau = 2.40\pm0.02$\,ps and $\omega_p/2\pi = 476.1 \pm 0.4$\,GHz, are in agreement with those obtained from fitting the rotation and ellipticity as a function of frequency. 

The residuals of $\phi$ and $\chi$ shown in Fig.\,\ref{fig:phi_chi_InSb_Hdep} indicate that the overall resolution for tracking small changes of  the rotation and ellipticity is about 1\,mrad at 442\,GHz.
A typical measurement time of an interferogram (Fig.\,\ref{fig:Interferogram_InSb}) of 600 data points is 30 minutes and the spacing of points near minima is as determined by the precision of the moving mirror stage.

\section{\label{Summary}Summary} 

We have developed a MOKE spectrometer designed for simultaneous measurement of the Kerr rotation angle and ellipticity using continuous wave sub-THz low-power radiation over a broad frequency range under normal incidence to the sample, eliminating the need for reference measurements. 
We demonstrated that the spectrometer is capable of measuring the frequency dependence of the polar Kerr effect response function with sub-milliradian accuracy. 

The MOKE spectrometer is well-suited for integration with magneto-optical systems that have direct optical access. 
The implemented interferometric technique can be adapted to other radiation sources across a broad frequency range, including shorter wavelengths, by using polarizing beam splitter cubes. 
The instrument’s versatility allows for the studies of magneto-optical effects in quantum materials such as unconventional superconductors, two-dimensional electron gas systems, quantum magnets, and systems exhibiting the anomalous Hall effect, at sub-Kelvin temperatures and in high magnetic fields.

\begin{acknowledgments}

The work was supported by the Estonian science agency grants PRG736, MOBJD1103 and by the European Research Council (ERC) under the European Unions Horizon 2020 research and innovation program Grant Agreement No. 885413. 
The work at Rutgers was supported by the National Science Foundation (NSF) Grant No. DMR-2105001.

\end{acknowledgments}

\section*{Author declarations}
\subsection*{Conflicts of interest}
The authors have no conflicts to disclose.
\subsection*{Author contributions}

GB conceived the project. 
UN, GB and TR invented the measurement method.
TR, GB and UN modelled the response function of the interferometer.
UN, GB and AGM built the interferometer.
UN created the LabVIEW code to run the Kerr interferometer and control the experiments.
All authors performed the experiments and wrote the manuscript.
{}
\section*{Data Availability Statement}
The data that support the findings of this study are available from the corresponding author upon reasonable request.

\section*{References}

\begin{thebibliography}{49}%
\makeatletter
\providecommand \@ifxundefined [1]{%
 \@ifx{#1\undefined}
}%
\providecommand \@ifnum [1]{%
 \ifnum #1\expandafter \@firstoftwo
 \else \expandafter \@secondoftwo
 \fi
}%
\providecommand \@ifx [1]{%
 \ifx #1\expandafter \@firstoftwo
 \else \expandafter \@secondoftwo
 \fi
}%
\providecommand \natexlab [1]{#1}%
\providecommand \enquote  [1]{``#1''}%
\providecommand \bibnamefont  [1]{#1}%
\providecommand \bibfnamefont [1]{#1}%
\providecommand \citenamefont [1]{#1}%
\providecommand \href@noop [0]{\@secondoftwo}%
\providecommand \href [0]{\begingroup \@sanitize@url \@href}%
\providecommand \@href[1]{\@@startlink{#1}\@@href}%
\providecommand \@@href[1]{\endgroup#1\@@endlink}%
\providecommand \@sanitize@url [0]{\catcode `\\12\catcode `\$12\catcode
  `\&12\catcode `\#12\catcode `\^12\catcode `\_12\catcode `\%12\relax}%
\providecommand \@@startlink[1]{}%
\providecommand \@@endlink[0]{}%
\providecommand \url  [0]{\begingroup\@sanitize@url \@url }%
\providecommand \@url [1]{\endgroup\@href {#1}{\urlprefix }}%
\providecommand \urlprefix  [0]{URL }%
\providecommand \Eprint [0]{\href }%
\providecommand \doibase [0]{https://doi.org/}%
\providecommand \selectlanguage [0]{\@gobble}%
\providecommand \bibinfo  [0]{\@secondoftwo}%
\providecommand \bibfield  [0]{\@secondoftwo}%
\providecommand \translation [1]{[#1]}%
\providecommand \BibitemOpen [0]{}%
\providecommand \bibitemStop [0]{}%
\providecommand \bibitemNoStop [0]{.\EOS\space}%
\providecommand \EOS [0]{\spacefactor3000\relax}%
\providecommand \BibitemShut  [1]{\csname bibitem#1\endcsname}%
\let\auto@bib@innerbib\@empty
\bibitem [{\citenamefont {Kimel}\ \emph {et~al.}(2022)\citenamefont {Kimel},
  \citenamefont {Zvezdin}, \citenamefont {Sharma}, \citenamefont {Shallcross},
  \citenamefont {de~Sousa}, \citenamefont {Garc{\'{\i}}a-Mart{\'{\i}}n},
  \citenamefont {Salvan}, \citenamefont {Hamrle}, \citenamefont {Stejskal},
  \citenamefont {McCord}, \citenamefont {Tacchi}, \citenamefont {Carlotti},
  \citenamefont {Gambardella}, \citenamefont {Salis}, \citenamefont
  {Münzenberg}, \citenamefont {Schultze}, \citenamefont {Temnov},
  \citenamefont {Bychkov}, \citenamefont {Kotov}, \citenamefont {Maccaferri},
  \citenamefont {Ignatyeva}, \citenamefont {Belotelov}, \citenamefont
  {Donnelly}, \citenamefont {Rodriguez}, \citenamefont {Matsuda}, \citenamefont
  {Ruchon}, \citenamefont {Fanciulli}, \citenamefont {Sacchi}, \citenamefont
  {Du}, \citenamefont {Wang}, \citenamefont {Armitage}, \citenamefont
  {Schubert}, \citenamefont {Darakchieva}, \citenamefont {Liu}, \citenamefont
  {Huang}, \citenamefont {Ding}, \citenamefont {Berger},\ and\ \citenamefont
  {Vavassori}}]{Kimel2022}%
  \BibitemOpen
  \bibfield  {author} {\bibinfo {author} {\bibfnamefont {A.}~\bibnamefont
  {Kimel}}, \bibinfo {author} {\bibfnamefont {A.}~\bibnamefont {Zvezdin}},
  \bibinfo {author} {\bibfnamefont {S.}~\bibnamefont {Sharma}}, \bibinfo
  {author} {\bibfnamefont {S.}~\bibnamefont {Shallcross}}, \bibinfo {author}
  {\bibfnamefont {N.}~\bibnamefont {de~Sousa}}, \bibinfo {author}
  {\bibfnamefont {A.}~\bibnamefont {Garc{\'{\i}}a-Mart{\'{\i}}n}}, \bibinfo
  {author} {\bibfnamefont {G.}~\bibnamefont {Salvan}}, \bibinfo {author}
  {\bibfnamefont {J.}~\bibnamefont {Hamrle}}, \bibinfo {author} {\bibfnamefont
  {O.}~\bibnamefont {Stejskal}}, \bibinfo {author} {\bibfnamefont
  {J.}~\bibnamefont {McCord}}, \bibinfo {author} {\bibfnamefont
  {S.}~\bibnamefont {Tacchi}}, \bibinfo {author} {\bibfnamefont
  {G.}~\bibnamefont {Carlotti}}, \bibinfo {author} {\bibfnamefont
  {P.}~\bibnamefont {Gambardella}}, \bibinfo {author} {\bibfnamefont
  {G.}~\bibnamefont {Salis}}, \bibinfo {author} {\bibfnamefont
  {M.}~\bibnamefont {Münzenberg}}, \bibinfo {author} {\bibfnamefont
  {M.}~\bibnamefont {Schultze}}, \bibinfo {author} {\bibfnamefont
  {V.}~\bibnamefont {Temnov}}, \bibinfo {author} {\bibfnamefont {I.~V.}\
  \bibnamefont {Bychkov}}, \bibinfo {author} {\bibfnamefont {L.~N.}\
  \bibnamefont {Kotov}}, \bibinfo {author} {\bibfnamefont {N.}~\bibnamefont
  {Maccaferri}}, \bibinfo {author} {\bibfnamefont {D.}~\bibnamefont
  {Ignatyeva}}, \bibinfo {author} {\bibfnamefont {V.}~\bibnamefont
  {Belotelov}}, \bibinfo {author} {\bibfnamefont {C.}~\bibnamefont {Donnelly}},
  \bibinfo {author} {\bibfnamefont {A.~H.}\ \bibnamefont {Rodriguez}}, \bibinfo
  {author} {\bibfnamefont {I.}~\bibnamefont {Matsuda}}, \bibinfo {author}
  {\bibfnamefont {T.}~\bibnamefont {Ruchon}}, \bibinfo {author} {\bibfnamefont
  {M.}~\bibnamefont {Fanciulli}}, \bibinfo {author} {\bibfnamefont
  {M.}~\bibnamefont {Sacchi}}, \bibinfo {author} {\bibfnamefont {C.~R.}\
  \bibnamefont {Du}}, \bibinfo {author} {\bibfnamefont {H.}~\bibnamefont
  {Wang}}, \bibinfo {author} {\bibfnamefont {N.~P.}\ \bibnamefont {Armitage}},
  \bibinfo {author} {\bibfnamefont {M.}~\bibnamefont {Schubert}}, \bibinfo
  {author} {\bibfnamefont {V.}~\bibnamefont {Darakchieva}}, \bibinfo {author}
  {\bibfnamefont {B.}~\bibnamefont {Liu}}, \bibinfo {author} {\bibfnamefont
  {Z.}~\bibnamefont {Huang}}, \bibinfo {author} {\bibfnamefont
  {B.}~\bibnamefont {Ding}}, \bibinfo {author} {\bibfnamefont {A.}~\bibnamefont
  {Berger}},\ and\ \bibinfo {author} {\bibfnamefont {P.}~\bibnamefont
  {Vavassori}},\ }\bibfield  {title} {\enquote {\bibinfo {title} {The 2022
  magneto-optics roadmap},}\ }\href {https://doi.org/10.1088/1361-6463/ac8da0}
  {\bibfield  {journal} {\bibinfo  {journal} {J. Phys. D: Appl. Phys.}\
  }\textbf {\bibinfo {volume} {55}},\ \bibinfo {pages} {463003} (\bibinfo
  {year} {2022})}\BibitemShut {NoStop}%
\bibitem [{\citenamefont {Levallois}\ \emph {et~al.}(2015)\citenamefont
  {Levallois}, \citenamefont {Nedoliuk}, \citenamefont {Crassee},\ and\
  \citenamefont {Kuzmenko}}]{Levallois2015}%
  \BibitemOpen
  \bibfield  {author} {\bibinfo {author} {\bibfnamefont {J.}~\bibnamefont
  {Levallois}}, \bibinfo {author} {\bibfnamefont {I.~O.}\ \bibnamefont
  {Nedoliuk}}, \bibinfo {author} {\bibfnamefont {I.}~\bibnamefont {Crassee}},\
  and\ \bibinfo {author} {\bibfnamefont {A.~B.}\ \bibnamefont {Kuzmenko}},\
  }\bibfield  {title} {\enquote {\bibinfo {title} {Magneto-optical
  {Kramers-Kronig} analysis},}\ }\href {https://doi.org/10.1063/1.4914846}
  {\bibfield  {journal} {\bibinfo  {journal} {Rev. Sci. Instrum.}\ }\textbf
  {\bibinfo {volume} {86}},\ \bibinfo {pages} {033906} (\bibinfo {year}
  {2015})}\BibitemShut {NoStop}%
\bibitem [{\citenamefont {Drew}\ and\ \citenamefont
  {Coleman}(1997)}]{Drew1997}%
  \BibitemOpen
  \bibfield  {author} {\bibinfo {author} {\bibfnamefont {H.~D.}\ \bibnamefont
  {Drew}}\ and\ \bibinfo {author} {\bibfnamefont {P.}~\bibnamefont {Coleman}},\
  }\bibfield  {title} {\enquote {\bibinfo {title} {Sum rule for the optical
  {Hall} angle},}\ }\href {https://doi.org/10.1103/physrevlett.78.1572}
  {\bibfield  {journal} {\bibinfo  {journal} {Phys. Rev. Lett.}\ }\textbf
  {\bibinfo {volume} {78}},\ \bibinfo {pages} {1572--1575} (\bibinfo {year}
  {1997})}\BibitemShut {NoStop}%
\bibitem [{\citenamefont {Cho}\ and\ \citenamefont {Kivelson}(2016)}]{Cho2016}%
  \BibitemOpen
  \bibfield  {author} {\bibinfo {author} {\bibfnamefont {W.}~\bibnamefont
  {Cho}}\ and\ \bibinfo {author} {\bibfnamefont {S.~A.}\ \bibnamefont
  {Kivelson}},\ }\bibfield  {title} {\enquote {\bibinfo {title} {{Necessity of
  Time-Reversal Symmetry Breaking for the Polar Kerr Effect in Linear
  Response}},}\ }\href {https://doi.org/10.1103/physrevlett.116.093903}
  {\bibfield  {journal} {\bibinfo  {journal} {Physical Review Letters}\
  }\textbf {\bibinfo {volume} {116}},\ \bibinfo {pages} {093903} (\bibinfo
  {year} {2016})}\BibitemShut {NoStop}%
\bibitem [{\citenamefont {Wang}\ \emph {et~al.}(2017)\citenamefont {Wang},
  \citenamefont {Berlinsky}, \citenamefont {Zwicknagl},\ and\ \citenamefont
  {Kallin}}]{Kallin2017}%
  \BibitemOpen
  \bibfield  {author} {\bibinfo {author} {\bibfnamefont {Z.}~\bibnamefont
  {Wang}}, \bibinfo {author} {\bibfnamefont {J.}~\bibnamefont {Berlinsky}},
  \bibinfo {author} {\bibfnamefont {G.}~\bibnamefont {Zwicknagl}},\ and\
  \bibinfo {author} {\bibfnamefont {C.}~\bibnamefont {Kallin}},\ }\bibfield
  {title} {\enquote {\bibinfo {title} {Intrinsic ac anomalous hall effect of
  nonsymmorphic chiral superconductors with an application to
  ${\mathrm{upt}}_{3}$},}\ }\href {https://doi.org/10.1103/PhysRevB.96.174511}
  {\bibfield  {journal} {\bibinfo  {journal} {Phys. Rev. B}\ }\textbf {\bibinfo
  {volume} {96}},\ \bibinfo {pages} {174511} (\bibinfo {year}
  {2017})}\BibitemShut {NoStop}%
\bibitem [{\citenamefont {Levenson-Falk}\ \emph {et~al.}(2018)\citenamefont
  {Levenson-Falk}, \citenamefont {Schemm}, \citenamefont {Aoki}, \citenamefont
  {Maple},\ and\ \citenamefont {Kapitulnik}}]{Kapitulnik2018}%
  \BibitemOpen
  \bibfield  {author} {\bibinfo {author} {\bibfnamefont {E.~M.}\ \bibnamefont
  {Levenson-Falk}}, \bibinfo {author} {\bibfnamefont {E.~R.}\ \bibnamefont
  {Schemm}}, \bibinfo {author} {\bibfnamefont {Y.}~\bibnamefont {Aoki}},
  \bibinfo {author} {\bibfnamefont {M.~B.}\ \bibnamefont {Maple}},\ and\
  \bibinfo {author} {\bibfnamefont {A.}~\bibnamefont {Kapitulnik}},\ }\bibfield
   {title} {\enquote {\bibinfo {title} {{Polar {Kerr} Effect from Time-Reversal
  Symmetry Breaking in the Heavy-Fermion Superconductor
  ${\mathrm{PrOs}}_{4}{\mathrm{Sb}}_{12}$}},}\ }\href
  {https://doi.org/10.1103/PhysRevLett.120.187004} {\bibfield  {journal}
  {\bibinfo  {journal} {Phys. Rev. Lett.}\ }\textbf {\bibinfo {volume} {120}},\
  \bibinfo {pages} {187004} (\bibinfo {year} {2018})}\BibitemShut {NoStop}%
\bibitem [{\citenamefont {Ikebe}\ \emph {et~al.}(2010)\citenamefont {Ikebe},
  \citenamefont {Morimoto}, \citenamefont {Masutomi}, \citenamefont {Okamoto},
  \citenamefont {Aoki},\ and\ \citenamefont {Shimano}}]{Ikebe2010}%
  \BibitemOpen
  \bibfield  {author} {\bibinfo {author} {\bibfnamefont {Y.}~\bibnamefont
  {Ikebe}}, \bibinfo {author} {\bibfnamefont {T.}~\bibnamefont {Morimoto}},
  \bibinfo {author} {\bibfnamefont {R.}~\bibnamefont {Masutomi}}, \bibinfo
  {author} {\bibfnamefont {T.}~\bibnamefont {Okamoto}}, \bibinfo {author}
  {\bibfnamefont {H.}~\bibnamefont {Aoki}},\ and\ \bibinfo {author}
  {\bibfnamefont {R.}~\bibnamefont {Shimano}},\ }\bibfield  {title} {\enquote
  {\bibinfo {title} {Optical {Hall} effect in the integer quantum {Hall}
  regime},}\ }\href {https://doi.org/10.1103/physrevlett.104.256802} {\bibfield
   {journal} {\bibinfo  {journal} {Phys. Rev. Lett.}\ }\textbf {\bibinfo
  {volume} {104}},\ \bibinfo {pages} {256802} (\bibinfo {year}
  {2010})}\BibitemShut {NoStop}%
\bibitem [{\citenamefont {Dziom}\ \emph {et~al.}(2019)\citenamefont {Dziom},
  \citenamefont {Shuvaev}, \citenamefont {Shchepetilnikov}, \citenamefont
  {MacFarland}, \citenamefont {Strasser},\ and\ \citenamefont
  {Pimenov}}]{Dziom2019}%
  \BibitemOpen
  \bibfield  {author} {\bibinfo {author} {\bibfnamefont {V.}~\bibnamefont
  {Dziom}}, \bibinfo {author} {\bibfnamefont {A.}~\bibnamefont {Shuvaev}},
  \bibinfo {author} {\bibfnamefont {A.~V.}\ \bibnamefont {Shchepetilnikov}},
  \bibinfo {author} {\bibfnamefont {D.}~\bibnamefont {MacFarland}}, \bibinfo
  {author} {\bibfnamefont {G.}~\bibnamefont {Strasser}},\ and\ \bibinfo
  {author} {\bibfnamefont {A.}~\bibnamefont {Pimenov}},\ }\bibfield  {title}
  {\enquote {\bibinfo {title} {High-frequency breakdown of the integer quantum
  {Hall} effect in {GaAs/AlGaAs} heterojunctions},}\ }\href
  {https://doi.org/10.1103/physrevb.99.045305} {\bibfield  {journal} {\bibinfo
  {journal} {Phys. Rev. B}\ }\textbf {\bibinfo {volume} {99}},\ \bibinfo
  {pages} {045305} (\bibinfo {year} {2019})}\BibitemShut {NoStop}%
\bibitem [{\citenamefont {Okada}\ \emph {et~al.}(2016)\citenamefont {Okada},
  \citenamefont {Takahashi}, \citenamefont {Mogi}, \citenamefont {Yoshimi},
  \citenamefont {Tsukazaki}, \citenamefont {Takahashi}, \citenamefont {Ogawa},
  \citenamefont {Kawasaki},\ and\ \citenamefont {Tokura}}]{Okada2016}%
  \BibitemOpen
  \bibfield  {author} {\bibinfo {author} {\bibfnamefont {K.~N.}\ \bibnamefont
  {Okada}}, \bibinfo {author} {\bibfnamefont {Y.}~\bibnamefont {Takahashi}},
  \bibinfo {author} {\bibfnamefont {M.}~\bibnamefont {Mogi}}, \bibinfo {author}
  {\bibfnamefont {R.}~\bibnamefont {Yoshimi}}, \bibinfo {author} {\bibfnamefont
  {A.}~\bibnamefont {Tsukazaki}}, \bibinfo {author} {\bibfnamefont {K.~S.}\
  \bibnamefont {Takahashi}}, \bibinfo {author} {\bibfnamefont {N.}~\bibnamefont
  {Ogawa}}, \bibinfo {author} {\bibfnamefont {M.}~\bibnamefont {Kawasaki}},\
  and\ \bibinfo {author} {\bibfnamefont {Y.}~\bibnamefont {Tokura}},\
  }\bibfield  {title} {\enquote {\bibinfo {title} {{Terahertz spectroscopy on
  Faraday and Kerr rotations in a quantum anomalous Hall state}},}\ }\href
  {https://doi.org/10.1038/ncomms12245} {\bibfield  {journal} {\bibinfo
  {journal} {Nature Communications}\ }\textbf {\bibinfo {volume} {7}},\
  \bibinfo {pages} {12245} (\bibinfo {year} {2016})}\BibitemShut {NoStop}%
\bibitem [{\citenamefont {Gusynin}, \citenamefont {Sharapov},\ and\
  \citenamefont {Carbotte}(2007)}]{Gusynin2007}%
  \BibitemOpen
  \bibfield  {author} {\bibinfo {author} {\bibfnamefont {V.~P.}\ \bibnamefont
  {Gusynin}}, \bibinfo {author} {\bibfnamefont {S.~G.}\ \bibnamefont
  {Sharapov}},\ and\ \bibinfo {author} {\bibfnamefont {J.~P.}\ \bibnamefont
  {Carbotte}},\ }\bibfield  {title} {\enquote {\bibinfo {title} {Sum rules for
  the optical and {Hall} conductivity in graphene},}\ }\href
  {https://doi.org/10.1103/physrevb.75.165407} {\bibfield  {journal} {\bibinfo
  {journal} {Phys. Rev. B}\ }\textbf {\bibinfo {volume} {75}},\ \bibinfo
  {pages} {165407} (\bibinfo {year} {2007})}\BibitemShut {NoStop}%
\bibitem [{\citenamefont {Yu}\ \emph {et~al.}(2010)\citenamefont {Yu},
  \citenamefont {Zhang}, \citenamefont {Zhang}, \citenamefont {Zhang},
  \citenamefont {Dai},\ and\ \citenamefont {Fang}}]{Fang2010}%
  \BibitemOpen
  \bibfield  {author} {\bibinfo {author} {\bibfnamefont {R.}~\bibnamefont
  {Yu}}, \bibinfo {author} {\bibfnamefont {W.}~\bibnamefont {Zhang}}, \bibinfo
  {author} {\bibfnamefont {H.-J.}\ \bibnamefont {Zhang}}, \bibinfo {author}
  {\bibfnamefont {S.-C.}\ \bibnamefont {Zhang}}, \bibinfo {author}
  {\bibfnamefont {X.}~\bibnamefont {Dai}},\ and\ \bibinfo {author}
  {\bibfnamefont {Z.}~\bibnamefont {Fang}},\ }\bibfield  {title} {\enquote
  {\bibinfo {title} {{Quantized Anomalous Hall Effect in Magnetic Topological
  Insulators}},}\ }\href {https://doi.org/10.1126/science.1187485} {\bibfield
  {journal} {\bibinfo  {journal} {Science}\ }\textbf {\bibinfo {volume}
  {329}},\ \bibinfo {pages} {61--64} (\bibinfo {year} {2010})},\ \Eprint
  {https://arxiv.org/abs/https://www.science.org/doi/pdf/10.1126/science.1187485}
  {https://www.science.org/doi/pdf/10.1126/science.1187485} \BibitemShut
  {NoStop}%
\bibitem [{\citenamefont {Chang}\ \emph {et~al.}(2013)\citenamefont {Chang},
  \citenamefont {Zhang}, \citenamefont {Feng}, \citenamefont {Shen},
  \citenamefont {Zhang}, \citenamefont {Guo}, \citenamefont {Li}, \citenamefont
  {Ou}, \citenamefont {Wei}, \citenamefont {Wang}, \citenamefont {Ji},
  \citenamefont {Feng}, \citenamefont {Ji}, \citenamefont {Chen}, \citenamefont
  {Jia}, \citenamefont {Dai}, \citenamefont {Fang}, \citenamefont {Zhang},
  \citenamefont {He}, \citenamefont {Wang}, \citenamefont {Lu}, \citenamefont
  {Ma},\ and\ \citenamefont {Xue}}]{Xue2013}%
  \BibitemOpen
  \bibfield  {author} {\bibinfo {author} {\bibfnamefont {C.-Z.}\ \bibnamefont
  {Chang}}, \bibinfo {author} {\bibfnamefont {J.}~\bibnamefont {Zhang}},
  \bibinfo {author} {\bibfnamefont {X.}~\bibnamefont {Feng}}, \bibinfo {author}
  {\bibfnamefont {J.}~\bibnamefont {Shen}}, \bibinfo {author} {\bibfnamefont
  {Z.}~\bibnamefont {Zhang}}, \bibinfo {author} {\bibfnamefont
  {M.}~\bibnamefont {Guo}}, \bibinfo {author} {\bibfnamefont {K.}~\bibnamefont
  {Li}}, \bibinfo {author} {\bibfnamefont {Y.}~\bibnamefont {Ou}}, \bibinfo
  {author} {\bibfnamefont {P.}~\bibnamefont {Wei}}, \bibinfo {author}
  {\bibfnamefont {L.-L.}\ \bibnamefont {Wang}}, \bibinfo {author}
  {\bibfnamefont {Z.-Q.}\ \bibnamefont {Ji}}, \bibinfo {author} {\bibfnamefont
  {Y.}~\bibnamefont {Feng}}, \bibinfo {author} {\bibfnamefont {S.}~\bibnamefont
  {Ji}}, \bibinfo {author} {\bibfnamefont {X.}~\bibnamefont {Chen}}, \bibinfo
  {author} {\bibfnamefont {J.}~\bibnamefont {Jia}}, \bibinfo {author}
  {\bibfnamefont {X.}~\bibnamefont {Dai}}, \bibinfo {author} {\bibfnamefont
  {Z.}~\bibnamefont {Fang}}, \bibinfo {author} {\bibfnamefont {S.-C.}\
  \bibnamefont {Zhang}}, \bibinfo {author} {\bibfnamefont {K.}~\bibnamefont
  {He}}, \bibinfo {author} {\bibfnamefont {Y.}~\bibnamefont {Wang}}, \bibinfo
  {author} {\bibfnamefont {L.}~\bibnamefont {Lu}}, \bibinfo {author}
  {\bibfnamefont {X.-C.}\ \bibnamefont {Ma}},\ and\ \bibinfo {author}
  {\bibfnamefont {Q.-K.}\ \bibnamefont {Xue}},\ }\bibfield  {title} {\enquote
  {\bibinfo {title} {{Experimental Observation of the Quantum Anomalous Hall
  Effect in a Magnetic Topological Insulator}},}\ }\href
  {https://doi.org/10.1126/science.1234414} {\bibfield  {journal} {\bibinfo
  {journal} {Science}\ }\textbf {\bibinfo {volume} {340}},\ \bibinfo {pages}
  {167--170} (\bibinfo {year} {2013})},\ \Eprint
  {https://arxiv.org/abs/https://www.science.org/doi/pdf/10.1126/science.1234414}
  {https://www.science.org/doi/pdf/10.1126/science.1234414} \BibitemShut
  {NoStop}%
\bibitem [{\citenamefont {Mak}\ \emph {et~al.}(2014)\citenamefont {Mak},
  \citenamefont {McGill}, \citenamefont {Park},\ and\ \citenamefont
  {McEuen}}]{McEuen2014}%
  \BibitemOpen
  \bibfield  {author} {\bibinfo {author} {\bibfnamefont {K.~F.}\ \bibnamefont
  {Mak}}, \bibinfo {author} {\bibfnamefont {K.~L.}\ \bibnamefont {McGill}},
  \bibinfo {author} {\bibfnamefont {J.}~\bibnamefont {Park}},\ and\ \bibinfo
  {author} {\bibfnamefont {P.~L.}\ \bibnamefont {McEuen}},\ }\bibfield  {title}
  {\enquote {\bibinfo {title} {{The valley Hall effect in MoS<sub>2</sub>
  transistors}},}\ }\href {https://doi.org/10.1126/science.1250140} {\bibfield
  {journal} {\bibinfo  {journal} {Science}\ }\textbf {\bibinfo {volume}
  {344}},\ \bibinfo {pages} {1489--1492} (\bibinfo {year} {2014})}\BibitemShut
  {NoStop}%
\bibitem [{\citenamefont {Kallin}\ and\ \citenamefont
  {Berlinsky}(2016)}]{Kallin2016}%
  \BibitemOpen
  \bibfield  {author} {\bibinfo {author} {\bibfnamefont {C.}~\bibnamefont
  {Kallin}}\ and\ \bibinfo {author} {\bibfnamefont {J.}~\bibnamefont
  {Berlinsky}},\ }\bibfield  {title} {\enquote {\bibinfo {title} {Chiral
  superconductors},}\ }\href {https://doi.org/10.1088/0034-4885/79/5/054502}
  {\bibfield  {journal} {\bibinfo  {journal} {Reports on Progress in Physics}\
  }\textbf {\bibinfo {volume} {79}},\ \bibinfo {pages} {054502} (\bibinfo
  {year} {2016})}\BibitemShut {NoStop}%
\bibitem [{\citenamefont {Sato}\ and\ \citenamefont {Ando}(2017)}]{Sato2017}%
  \BibitemOpen
  \bibfield  {author} {\bibinfo {author} {\bibfnamefont {M.}~\bibnamefont
  {Sato}}\ and\ \bibinfo {author} {\bibfnamefont {Y.}~\bibnamefont {Ando}},\
  }\bibfield  {title} {\enquote {\bibinfo {title} {Topological superconductors:
  a review},}\ }\href {https://doi.org/10.1088/1361-6633/aa6ac7} {\bibfield
  {journal} {\bibinfo  {journal} {Rep. Prog. Phys.}\ }\textbf {\bibinfo
  {volume} {80}},\ \bibinfo {pages} {076501} (\bibinfo {year}
  {2017})}\BibitemShut {NoStop}%
\bibitem [{\citenamefont {Ghosh}\ \emph {et~al.}(2020)\citenamefont {Ghosh},
  \citenamefont {Smidman}, \citenamefont {Shang}, \citenamefont {Annett},
  \citenamefont {Hillier}, \citenamefont {Quintanilla},\ and\ \citenamefont
  {Yuan}}]{Ghosh2020}%
  \BibitemOpen
  \bibfield  {author} {\bibinfo {author} {\bibfnamefont {S.~K.}\ \bibnamefont
  {Ghosh}}, \bibinfo {author} {\bibfnamefont {M.}~\bibnamefont {Smidman}},
  \bibinfo {author} {\bibfnamefont {T.}~\bibnamefont {Shang}}, \bibinfo
  {author} {\bibfnamefont {J.~F.}\ \bibnamefont {Annett}}, \bibinfo {author}
  {\bibfnamefont {A.~D.}\ \bibnamefont {Hillier}}, \bibinfo {author}
  {\bibfnamefont {J.}~\bibnamefont {Quintanilla}},\ and\ \bibinfo {author}
  {\bibfnamefont {H.}~\bibnamefont {Yuan}},\ }\bibfield  {title} {\enquote
  {\bibinfo {title} {Recent progress on superconductors with time-reversal
  symmetry breaking},}\ }\href {https://doi.org/10.1088/1361-648x/abaa06}
  {\bibfield  {journal} {\bibinfo  {journal} {Journal of Physics: Condensed
  Matter}\ }\textbf {\bibinfo {volume} {33}},\ \bibinfo {pages} {033001}
  (\bibinfo {year} {2020})}\BibitemShut {NoStop}%
\bibitem [{\citenamefont {Kapitulnik}\ \emph {et~al.}(2009)\citenamefont
  {Kapitulnik}, \citenamefont {Xia}, \citenamefont {Schemm},\ and\
  \citenamefont {Palevski}}]{Kapitulnik2009}%
  \BibitemOpen
  \bibfield  {author} {\bibinfo {author} {\bibfnamefont {A.}~\bibnamefont
  {Kapitulnik}}, \bibinfo {author} {\bibfnamefont {J.}~\bibnamefont {Xia}},
  \bibinfo {author} {\bibfnamefont {E.}~\bibnamefont {Schemm}},\ and\ \bibinfo
  {author} {\bibfnamefont {A.}~\bibnamefont {Palevski}},\ }\bibfield  {title}
  {\enquote {\bibinfo {title} {Polar {Kerr} effect as probe for time-reversal
  symmetry breaking in unconventional superconductors},}\ }\href
  {https://doi.org/10.1088/1367-2630/11/5/055060} {\bibfield  {journal}
  {\bibinfo  {journal} {New Journal of Physics}\ }\textbf {\bibinfo {volume}
  {11}},\ \bibinfo {pages} {055060} (\bibinfo {year} {2009})}\BibitemShut
  {NoStop}%
\bibitem [{\citenamefont {Brydon}\ \emph {et~al.}(2019)\citenamefont {Brydon},
  \citenamefont {Abergel}, \citenamefont {Agterberg},\ and\ \citenamefont
  {Yakovenko}}]{Yakovenko2019PRX}%
  \BibitemOpen
  \bibfield  {author} {\bibinfo {author} {\bibfnamefont {P.}~\bibnamefont
  {Brydon}}, \bibinfo {author} {\bibfnamefont {D.}~\bibnamefont {Abergel}},
  \bibinfo {author} {\bibfnamefont {D.}~\bibnamefont {Agterberg}},\ and\
  \bibinfo {author} {\bibfnamefont {V.~M.}\ \bibnamefont {Yakovenko}},\
  }\bibfield  {title} {\enquote {\bibinfo {title} {Loop currents and anomalous
  hall effect from time-reversal symmetry-breaking superconductivity on the
  honeycomb lattice},}\ }\href {https://doi.org/10.1103/physrevx.9.031025}
  {\bibfield  {journal} {\bibinfo  {journal} {Physical Review X}\ }\textbf
  {\bibinfo {volume} {9}},\ \bibinfo {pages} {031025} (\bibinfo {year}
  {2019})}\BibitemShut {NoStop}%
\bibitem [{\citenamefont {Kapitulnik}(2015)}]{Kapitulnik2015notes}%
  \BibitemOpen
  \bibfield  {author} {\bibinfo {author} {\bibfnamefont {A.}~\bibnamefont
  {Kapitulnik}},\ }\bibfield  {title} {\enquote {\bibinfo {title} {Notes on
  constraints for the observation of polar kerr effect in complex materials},}\
  }\href {https://doi.org/https://doi.org/10.1016/j.physb.2014.11.059}
  {\bibfield  {journal} {\bibinfo  {journal} {Phys. B Condens. Matter}\
  }\textbf {\bibinfo {volume} {460}},\ \bibinfo {pages} {151--158} (\bibinfo
  {year} {2015})},\ \bibinfo {note} {special Issue on Electronic Crystals
  (ECRYS-2014)}\BibitemShut {NoStop}%
\bibitem [{\citenamefont {Schemm}\ \emph {et~al.}(2014)\citenamefont {Schemm},
  \citenamefont {Gannon}, \citenamefont {Wishne}, \citenamefont {Halperin},\
  and\ \citenamefont {Kapitulnik}}]{Kapitulnik2014}%
  \BibitemOpen
  \bibfield  {author} {\bibinfo {author} {\bibfnamefont {E.~R.}\ \bibnamefont
  {Schemm}}, \bibinfo {author} {\bibfnamefont {W.~J.}\ \bibnamefont {Gannon}},
  \bibinfo {author} {\bibfnamefont {C.~M.}\ \bibnamefont {Wishne}}, \bibinfo
  {author} {\bibfnamefont {W.~P.}\ \bibnamefont {Halperin}},\ and\ \bibinfo
  {author} {\bibfnamefont {A.}~\bibnamefont {Kapitulnik}},\ }\bibfield  {title}
  {\enquote {\bibinfo {title} {{Observation of broken time-reversal symmetry in
  the heavy-fermion superconductor UPt$_3$}},}\ }\href
  {https://doi.org/10.1126/science.1248552} {\bibfield  {journal} {\bibinfo
  {journal} {Science}\ }\textbf {\bibinfo {volume} {345}},\ \bibinfo {pages}
  {190--193} (\bibinfo {year} {2014})},\ \Eprint
  {https://arxiv.org/abs/https://science.sciencemag.org/content/345/6193/190.full.pdf}
  {https://science.sciencemag.org/content/345/6193/190.full.pdf} \BibitemShut
  {NoStop}%
\bibitem [{\citenamefont {Schemm}\ \emph {et~al.}(2015)\citenamefont {Schemm},
  \citenamefont {Baumbach}, \citenamefont {Tobash}, \citenamefont {Ronning},
  \citenamefont {Bauer},\ and\ \citenamefont {Kapitulnik}}]{Kapitulnik2015u}%
  \BibitemOpen
  \bibfield  {author} {\bibinfo {author} {\bibfnamefont {E.~R.}\ \bibnamefont
  {Schemm}}, \bibinfo {author} {\bibfnamefont {R.~E.}\ \bibnamefont
  {Baumbach}}, \bibinfo {author} {\bibfnamefont {P.~H.}\ \bibnamefont
  {Tobash}}, \bibinfo {author} {\bibfnamefont {F.}~\bibnamefont {Ronning}},
  \bibinfo {author} {\bibfnamefont {E.~D.}\ \bibnamefont {Bauer}},\ and\
  \bibinfo {author} {\bibfnamefont {A.}~\bibnamefont {Kapitulnik}},\ }\bibfield
   {title} {\enquote {\bibinfo {title} {{Evidence for broken time-reversal
  symmetry in the superconducting phase of URu$_2$Si$_2$}},}\ }\href
  {https://doi.org/10.1103/PhysRevB.91.140506} {\bibfield  {journal} {\bibinfo
  {journal} {Phys. Rev. B - Condens. Matter Mater. Phys.}\ }\textbf {\bibinfo
  {volume} {91}},\ \bibinfo {pages} {140506} (\bibinfo {year} {2015})},\
  \Eprint {https://arxiv.org/abs/1410.1479} {arXiv:1410.1479} \BibitemShut
  {NoStop}%
\bibitem [{\citenamefont {Hayes}\ \emph {et~al.}(2021)\citenamefont {Hayes},
  \citenamefont {Wei}, \citenamefont {Metz}, \citenamefont {Zhang},
  \citenamefont {Eo}, \citenamefont {Ran}, \citenamefont {Saha}, \citenamefont
  {Collini}, \citenamefont {Butch}, \citenamefont {Agterberg}, \citenamefont
  {Kapitulnik},\ and\ \citenamefont {Paglione}}]{Hayes2021}%
  \BibitemOpen
  \bibfield  {author} {\bibinfo {author} {\bibfnamefont {I.~M.}\ \bibnamefont
  {Hayes}}, \bibinfo {author} {\bibfnamefont {D.~S.}\ \bibnamefont {Wei}},
  \bibinfo {author} {\bibfnamefont {T.}~\bibnamefont {Metz}}, \bibinfo {author}
  {\bibfnamefont {J.}~\bibnamefont {Zhang}}, \bibinfo {author} {\bibfnamefont
  {Y.~S.}\ \bibnamefont {Eo}}, \bibinfo {author} {\bibfnamefont
  {S.}~\bibnamefont {Ran}}, \bibinfo {author} {\bibfnamefont {S.~R.}\
  \bibnamefont {Saha}}, \bibinfo {author} {\bibfnamefont {J.}~\bibnamefont
  {Collini}}, \bibinfo {author} {\bibfnamefont {N.~P.}\ \bibnamefont {Butch}},
  \bibinfo {author} {\bibfnamefont {D.~F.}\ \bibnamefont {Agterberg}}, \bibinfo
  {author} {\bibfnamefont {A.}~\bibnamefont {Kapitulnik}},\ and\ \bibinfo
  {author} {\bibfnamefont {J.}~\bibnamefont {Paglione}},\ }\bibfield  {title}
  {\enquote {\bibinfo {title} {Multicomponent superconducting order parameter
  in {UTe}$_2$},}\ }\href {https://doi.org/10.1126/science.abb0272} {\bibfield
  {journal} {\bibinfo  {journal} {Science}\ }\textbf {\bibinfo {volume}
  {373}},\ \bibinfo {pages} {797--801} (\bibinfo {year} {2021})}\BibitemShut
  {NoStop}%
\bibitem [{\citenamefont {Farhang}\ \emph {et~al.}(2023)\citenamefont
  {Farhang}, \citenamefont {Zaki}, \citenamefont {Wang}, \citenamefont {Gu},
  \citenamefont {Johnson},\ and\ \citenamefont {Xia}}]{Farhang2023}%
  \BibitemOpen
  \bibfield  {author} {\bibinfo {author} {\bibfnamefont {C.}~\bibnamefont
  {Farhang}}, \bibinfo {author} {\bibfnamefont {N.}~\bibnamefont {Zaki}},
  \bibinfo {author} {\bibfnamefont {J.}~\bibnamefont {Wang}}, \bibinfo {author}
  {\bibfnamefont {G.}~\bibnamefont {Gu}}, \bibinfo {author} {\bibfnamefont
  {P.~D.}\ \bibnamefont {Johnson}},\ and\ \bibinfo {author} {\bibfnamefont
  {J.}~\bibnamefont {Xia}},\ }\bibfield  {title} {\enquote {\bibinfo {title}
  {Revealing the origin of time-reversal symmetry breaking in {Fe}-chalcogenide
  superconductor {FeTe$_{1-x}$Se$_x$}},}\ }\href
  {https://doi.org/10.1103/physrevlett.130.046702} {\bibfield  {journal}
  {\bibinfo  {journal} {Phys. Rev. Lett.}\ }\textbf {\bibinfo {volume} {130}},\
  \bibinfo {pages} {046702} (\bibinfo {year} {2023})}\BibitemShut {NoStop}%
\bibitem [{\citenamefont {Ajeesh}\ \emph {et~al.}(2023)\citenamefont {Ajeesh},
  \citenamefont {Bordelon}, \citenamefont {Girod}, \citenamefont {Mishra},
  \citenamefont {Ronning}, \citenamefont {Bauer}, \citenamefont {Maiorov},
  \citenamefont {Thompson}, \citenamefont {Rosa},\ and\ \citenamefont
  {Thomas}}]{UTe2_Rosa2023}%
  \BibitemOpen
  \bibfield  {author} {\bibinfo {author} {\bibfnamefont {M.~O.}\ \bibnamefont
  {Ajeesh}}, \bibinfo {author} {\bibfnamefont {M.}~\bibnamefont {Bordelon}},
  \bibinfo {author} {\bibfnamefont {C.}~\bibnamefont {Girod}}, \bibinfo
  {author} {\bibfnamefont {S.}~\bibnamefont {Mishra}}, \bibinfo {author}
  {\bibfnamefont {F.}~\bibnamefont {Ronning}}, \bibinfo {author} {\bibfnamefont
  {E.~D.}\ \bibnamefont {Bauer}}, \bibinfo {author} {\bibfnamefont
  {B.}~\bibnamefont {Maiorov}}, \bibinfo {author} {\bibfnamefont {J.~D.}\
  \bibnamefont {Thompson}}, \bibinfo {author} {\bibfnamefont {P.~F.~S.}\
  \bibnamefont {Rosa}},\ and\ \bibinfo {author} {\bibfnamefont {S.~M.}\
  \bibnamefont {Thomas}},\ }\bibfield  {title} {\enquote {\bibinfo {title}
  {{Fate of Time-Reversal Symmetry Breaking in ${\mathrm{UTe}}_{2}$}},}\ }\href
  {https://doi.org/10.1103/PhysRevX.13.041019} {\bibfield  {journal} {\bibinfo
  {journal} {Phys. Rev. X}\ }\textbf {\bibinfo {volume} {13}},\ \bibinfo
  {pages} {041019} (\bibinfo {year} {2023})}\BibitemShut {NoStop}%
\bibitem [{\citenamefont {Azari}\ \emph {et~al.}(2023)\citenamefont {Azari},
  \citenamefont {Yakovlev}, \citenamefont {Rye}, \citenamefont {Dunsiger},
  \citenamefont {Sundar}, \citenamefont {Bordelon}, \citenamefont {Thomas},
  \citenamefont {Thompson}, \citenamefont {Rosa},\ and\ \citenamefont
  {Sonier}}]{UTe2_RosaPRL2023}%
  \BibitemOpen
  \bibfield  {author} {\bibinfo {author} {\bibfnamefont {N.}~\bibnamefont
  {Azari}}, \bibinfo {author} {\bibfnamefont {M.}~\bibnamefont {Yakovlev}},
  \bibinfo {author} {\bibfnamefont {N.}~\bibnamefont {Rye}}, \bibinfo {author}
  {\bibfnamefont {S.~R.}\ \bibnamefont {Dunsiger}}, \bibinfo {author}
  {\bibfnamefont {S.}~\bibnamefont {Sundar}}, \bibinfo {author} {\bibfnamefont
  {M.~M.}\ \bibnamefont {Bordelon}}, \bibinfo {author} {\bibfnamefont {S.~M.}\
  \bibnamefont {Thomas}}, \bibinfo {author} {\bibfnamefont {J.~D.}\
  \bibnamefont {Thompson}}, \bibinfo {author} {\bibfnamefont {P.~F.~S.}\
  \bibnamefont {Rosa}},\ and\ \bibinfo {author} {\bibfnamefont {J.~E.}\
  \bibnamefont {Sonier}},\ }\bibfield  {title} {\enquote {\bibinfo {title}
  {Absence of spontaneous magnetic fields due to time-reversal symmetry
  breaking in bulk superconducting ${\mathrm{ute}}_{2}$},}\ }\href
  {https://doi.org/10.1103/PhysRevLett.131.226504} {\bibfield  {journal}
  {\bibinfo  {journal} {Phys. Rev. Lett.}\ }\textbf {\bibinfo {volume} {131}},\
  \bibinfo {pages} {226504} (\bibinfo {year} {2023})}\BibitemShut {NoStop}%
\bibitem [{\citenamefont {Shimano}\ \emph {et~al.}(2002)\citenamefont
  {Shimano}, \citenamefont {Ino}, \citenamefont {Svirko},\ and\ \citenamefont
  {Kuwata-Gonokami}}]{Shimano2002}%
  \BibitemOpen
  \bibfield  {author} {\bibinfo {author} {\bibfnamefont {R.}~\bibnamefont
  {Shimano}}, \bibinfo {author} {\bibfnamefont {Y.}~\bibnamefont {Ino}},
  \bibinfo {author} {\bibfnamefont {Y.~P.}\ \bibnamefont {Svirko}},\ and\
  \bibinfo {author} {\bibfnamefont {M.}~\bibnamefont {Kuwata-Gonokami}},\
  }\bibfield  {title} {\enquote {\bibinfo {title} {Terahertz frequency {Hall}
  measurement by magneto-optical {Kerr} spectroscopy in {InAs}},}\ }\href
  {https://doi.org/10.1063/1.1492319} {\bibfield  {journal} {\bibinfo
  {journal} {Applied Physics Letters}\ }\textbf {\bibinfo {volume} {81}},\
  \bibinfo {pages} {199--201} (\bibinfo {year} {2002})}\BibitemShut {NoStop}%
\bibitem [{\citenamefont {Ino}\ \emph {et~al.}(2004)\citenamefont {Ino},
  \citenamefont {Shimano}, \citenamefont {Svirko},\ and\ \citenamefont
  {Kuwata-Gonokami}}]{Ino2004}%
  \BibitemOpen
  \bibfield  {author} {\bibinfo {author} {\bibfnamefont {Y.}~\bibnamefont
  {Ino}}, \bibinfo {author} {\bibfnamefont {R.}~\bibnamefont {Shimano}},
  \bibinfo {author} {\bibfnamefont {Y.}~\bibnamefont {Svirko}},\ and\ \bibinfo
  {author} {\bibfnamefont {M.}~\bibnamefont {Kuwata-Gonokami}},\ }\bibfield
  {title} {\enquote {\bibinfo {title} {Terahertz time domain magneto-optical
  ellipsometry in reflection geometry},}\ }\href
  {https://doi.org/10.1103/PhysRevB.70.155101} {\bibfield  {journal} {\bibinfo
  {journal} {Phys. Rev. B}\ }\textbf {\bibinfo {volume} {70}},\ \bibinfo
  {pages} {155101} (\bibinfo {year} {2004})}\BibitemShut {NoStop}%
\bibitem [{\citenamefont {Jenkins}, \citenamefont {Schmadel},\ and\
  \citenamefont {Drew}(2010)}]{Jenkins2010rsi}%
  \BibitemOpen
  \bibfield  {author} {\bibinfo {author} {\bibfnamefont {G.~S.}\ \bibnamefont
  {Jenkins}}, \bibinfo {author} {\bibfnamefont {D.~C.}\ \bibnamefont
  {Schmadel}},\ and\ \bibinfo {author} {\bibfnamefont {H.~D.}\ \bibnamefont
  {Drew}},\ }\bibfield  {title} {\enquote {\bibinfo {title} {Simultaneous
  measurement of circular dichroism and {Faraday} rotation at terahertz
  frequencies utilizing electric field sensitive detection via polarization
  modulation},}\ }\href {https://doi.org/10.1063/1.3480554} {\bibfield
  {journal} {\bibinfo  {journal} {Rev. Sci. Instrum.}\ }\textbf {\bibinfo
  {volume} {81}},\ \bibinfo {pages} {083903} (\bibinfo {year}
  {2010})}\BibitemShut {NoStop}%
\bibitem [{\citenamefont {Jenkins}\ \emph {et~al.}(2010)\citenamefont
  {Jenkins}, \citenamefont {Sushkov}, \citenamefont {Schmadel}, \citenamefont
  {Butch}, \citenamefont {Syers}, \citenamefont {Paglione},\ and\ \citenamefont
  {Drew}}]{Jenkins2010prb}%
  \BibitemOpen
  \bibfield  {author} {\bibinfo {author} {\bibfnamefont {G.~S.}\ \bibnamefont
  {Jenkins}}, \bibinfo {author} {\bibfnamefont {A.~B.}\ \bibnamefont
  {Sushkov}}, \bibinfo {author} {\bibfnamefont {D.~C.}\ \bibnamefont
  {Schmadel}}, \bibinfo {author} {\bibfnamefont {N.~P.}\ \bibnamefont {Butch}},
  \bibinfo {author} {\bibfnamefont {P.}~\bibnamefont {Syers}}, \bibinfo
  {author} {\bibfnamefont {J.}~\bibnamefont {Paglione}},\ and\ \bibinfo
  {author} {\bibfnamefont {H.~D.}\ \bibnamefont {Drew}},\ }\bibfield  {title}
  {\enquote {\bibinfo {title} {Terahertz kerr and reflectivity measurements on
  the topological insulator ${\text{bi}}_{2}{\text{se}}_{3}$},}\ }\href
  {https://doi.org/10.1103/PhysRevB.82.125120} {\bibfield  {journal} {\bibinfo
  {journal} {PRB}\ }\textbf {\bibinfo {volume} {82}},\ \bibinfo {pages}
  {125120} (\bibinfo {year} {2010})}\BibitemShut {NoStop}%
\bibitem [{\citenamefont {Tagay}, \citenamefont {Romero},\ and\ \citenamefont
  {Armitage}(2024)}]{Tagay2024}%
  \BibitemOpen
  \bibfield  {author} {\bibinfo {author} {\bibfnamefont {Z.}~\bibnamefont
  {Tagay}}, \bibinfo {author} {\bibfnamefont {R.}~\bibnamefont {Romero}},\ and\
  \bibinfo {author} {\bibfnamefont {N.~P.}\ \bibnamefont {Armitage}},\
  }\bibfield  {title} {\enquote {\bibinfo {title} {High-precision measurements
  of terahertz polarization states with a fiber coupled time-domain {THz}
  spectrometer},}\ }\href {https://doi.org/10.1364/OE.516736} {\bibfield
  {journal} {\bibinfo  {journal} {Optics Express}\ }\textbf {\bibinfo {volume}
  {32}},\ \bibinfo {pages} {15946} (\bibinfo {year} {2024})}\BibitemShut
  {NoStop}%
\bibitem [{\citenamefont {Neshat}\ and\ \citenamefont
  {Armitage}(2013)}]{Neshat2013}%
  \BibitemOpen
  \bibfield  {author} {\bibinfo {author} {\bibfnamefont {M.}~\bibnamefont
  {Neshat}}\ and\ \bibinfo {author} {\bibfnamefont {N.}~\bibnamefont
  {Armitage}},\ }\bibfield  {title} {\enquote {\bibinfo {title} {Developments
  in {THz} range ellipsometry},}\ }\href
  {https://doi.org/https://doi.org/10.1007/s10762-013-9984-4} {\bibfield
  {journal} {\bibinfo  {journal} {J Infrared Milli Terahz Waves}\ }\textbf
  {\bibinfo {volume} {34}},\ \bibinfo {pages} {682--708} (\bibinfo {year}
  {2013})}\BibitemShut {NoStop}%
\bibitem [{\citenamefont {Born}\ and\ \citenamefont {Wolf}(1999)}]{BornWolf}%
  \BibitemOpen
  \bibfield  {author} {\bibinfo {author} {\bibfnamefont {M.}~\bibnamefont
  {Born}}\ and\ \bibinfo {author} {\bibfnamefont {E.}~\bibnamefont {Wolf}},\
  }\href@noop {} {\emph {\bibinfo {title} {Principles of Optics}}}\ (\bibinfo
  {publisher} {Cambridge University Press},\ \bibinfo {year}
  {1999})\BibitemShut {NoStop}%
\bibitem [{\citenamefont {Argyres}(1955)}]{Argyres1955}%
  \BibitemOpen
  \bibfield  {author} {\bibinfo {author} {\bibfnamefont {P.~N.}\ \bibnamefont
  {Argyres}},\ }\bibfield  {title} {\enquote {\bibinfo {title} {Theory of the
  {Faraday} and {Kerr} effects in ferromagnetics},}\ }\href
  {https://doi.org/10.1103/PhysRev.97.334} {\bibfield  {journal} {\bibinfo
  {journal} {Phys. Rev.}\ }\textbf {\bibinfo {volume} {97}},\ \bibinfo {pages}
  {334--345} (\bibinfo {year} {1955})}\BibitemShut {NoStop}%
\bibitem [{\citenamefont {Martin}\ and\ \citenamefont
  {Puplett}(1969)}]{Martin1970}%
  \BibitemOpen
  \bibfield  {author} {\bibinfo {author} {\bibfnamefont {D.}~\bibnamefont
  {Martin}}\ and\ \bibinfo {author} {\bibfnamefont {E.}~\bibnamefont
  {Puplett}},\ }\bibfield  {title} {\enquote {\bibinfo {title} {Polarised
  interferometric spectrometry for the millimetre and submillimetre
  spectrum},}\ }\href
  {https://doi.org/http://dx.doi.org/10.1016/0020-0891(70)90006-0} {\bibfield
  {journal} {\bibinfo  {journal} {Infrared Phys.}\ }\textbf {\bibinfo {volume}
  {10}},\ \bibinfo {pages} {105 -- 109} (\bibinfo {year} {1969})}\BibitemShut
  {NoStop}%
\bibitem [{Note1()}]{Note1}%
  \BibitemOpen
  \bibinfo {note} {Interferometer arms are the ones with roof mirrors with the
  beamsplitter in between. The modified MP interferometer has two symmetric
  ports on the other side of the beamsplitter: input and output.}\BibitemShut
  {Stop}%
\bibitem [{Note2()}]{Note2}%
  \BibitemOpen
  \bibinfo {note} {The latter can be consider as the interferogram phase
  fluctuation noise.}\BibitemShut {Stop}%
\bibitem [{Note3()}]{Note3}%
  \BibitemOpen
  \bibinfo {note} {This accuracy can be further improved by implementing an
  amplitude modulation measurement schema.}\BibitemShut {Stop}%
\bibitem [{Note4()}]{Note4}%
  \BibitemOpen
  \bibinfo {note} {Stages with air bearing and interferometric position readout
  provide better positioning accuracy.}\BibitemShut {Stop}%
\bibitem [{Note5()}]{Note5}%
  \BibitemOpen
  \bibinfo {note} {Additionally, the sign of $\Phi $ could be obtained from the
  behaviour of the phase of the detected signal. We have not used the phase
  information so far, because of too large overall phase drifts that are caused
  by THz frequency and optical setup instabilities, including the lasers and
  fibers that connect the lasers to the photomixers.}\BibitemShut {Stop}%
\bibitem [{\citenamefont
  {https://www.toptica.com/products/terahertz-systems/frequency
  domain/terascan/}()}]{TopticaCW}%
  \BibitemOpen
  \bibfield  {author} {\bibinfo {author} {\bibnamefont
  {https://www.toptica.com/products/terahertz-systems/frequency
  domain/terascan/}}\ }\href@noop {} {}\BibitemShut {NoStop}%
\bibitem [{\citenamefont {Roggenbuck}\ \emph {et~al.}(2010)\citenamefont
  {Roggenbuck}, \citenamefont {{Schmitz}}, \citenamefont {Deninger},
  \citenamefont {Mayorga}, \citenamefont {Hemberger}, \citenamefont
  {G{\"u}sten},\ and\ \citenamefont {Gr{\"u}ninger}}]{Roggenbuck2010}%
  \BibitemOpen
  \bibfield  {author} {\bibinfo {author} {\bibfnamefont {A.}~\bibnamefont
  {Roggenbuck}}, \bibinfo {author} {\bibfnamefont {H.}~\bibnamefont
  {{Schmitz}}}, \bibinfo {author} {\bibfnamefont {A.}~\bibnamefont {Deninger}},
  \bibinfo {author} {\bibfnamefont {I.~C.}\ \bibnamefont {Mayorga}}, \bibinfo
  {author} {\bibfnamefont {J.}~\bibnamefont {Hemberger}}, \bibinfo {author}
  {\bibfnamefont {R.}~\bibnamefont {G{\"u}sten}},\ and\ \bibinfo {author}
  {\bibfnamefont {M.}~\bibnamefont {Gr{\"u}ninger}},\ }\bibfield  {title}
  {\enquote {\bibinfo {title} {Coherent broadband continuous-wave terahertz
  spectroscopy on solid-state samples},}\ }\href
  {http://stacks.iop.org/1367-2630/12/i=4/a=043017} {\bibfield  {journal}
  {\bibinfo  {journal} {New J. Phys.}\ }\textbf {\bibinfo {volume} {12}},\
  \bibinfo {pages} {043017} (\bibinfo {year} {2010})}\BibitemShut {NoStop}%
\bibitem [{\citenamefont {Roggenbuck}\ \emph {et~al.}(2013)\citenamefont
  {Roggenbuck}, \citenamefont {Langenbach}, \citenamefont {Thirunavukkuarasu},
  \citenamefont {Schmitz}, \citenamefont {Deninger}, \citenamefont {Mayorga},
  \citenamefont {G{\"u}sten}, \citenamefont {Hemberger},\ and\ \citenamefont
  {Gr{\"u}ninger}}]{Roggenbuck2013}%
  \BibitemOpen
  \bibfield  {author} {\bibinfo {author} {\bibfnamefont {A.}~\bibnamefont
  {Roggenbuck}}, \bibinfo {author} {\bibfnamefont {M.}~\bibnamefont
  {Langenbach}}, \bibinfo {author} {\bibfnamefont {K.}~\bibnamefont
  {Thirunavukkuarasu}}, \bibinfo {author} {\bibfnamefont {H.}~\bibnamefont
  {Schmitz}}, \bibinfo {author} {\bibfnamefont {A.}~\bibnamefont {Deninger}},
  \bibinfo {author} {\bibfnamefont {I.~C.}\ \bibnamefont {Mayorga}}, \bibinfo
  {author} {\bibfnamefont {R.}~\bibnamefont {G{\"u}sten}}, \bibinfo {author}
  {\bibfnamefont {J.}~\bibnamefont {Hemberger}},\ and\ \bibinfo {author}
  {\bibfnamefont {M.}~\bibnamefont {Gr{\"u}ninger}},\ }\bibfield  {title}
  {\enquote {\bibinfo {title} {Enhancing the stability of a continuous-wave
  terahertz system by photocurrent normalization},}\ }\href
  {https://doi.org/10.1364/JOSAB.30.001397} {\bibfield  {journal} {\bibinfo
  {journal} {J. Opt. Soc. Am. B}\ }\textbf {\bibinfo {volume} {30}},\ \bibinfo
  {pages} {1397--1401} (\bibinfo {year} {2013})}\BibitemShut {NoStop}%
\bibitem [{\citenamefont {Roggenbuck}\ \emph {et~al.}(2012)\citenamefont
  {Roggenbuck}, \citenamefont {Thirunavukkuarasu}, \citenamefont {Schmitz},
  \citenamefont {Marx}, \citenamefont {Deninger}, \citenamefont {Mayorga},
  \citenamefont {G\"{u}sten}, \citenamefont {Hemberger},\ and\ \citenamefont
  {Gr\"{u}ninger}}]{Roggenbuck2012stretcher}%
  \BibitemOpen
  \bibfield  {author} {\bibinfo {author} {\bibfnamefont {A.}~\bibnamefont
  {Roggenbuck}}, \bibinfo {author} {\bibfnamefont {K.}~\bibnamefont
  {Thirunavukkuarasu}}, \bibinfo {author} {\bibfnamefont {H.}~\bibnamefont
  {Schmitz}}, \bibinfo {author} {\bibfnamefont {J.}~\bibnamefont {Marx}},
  \bibinfo {author} {\bibfnamefont {A.}~\bibnamefont {Deninger}}, \bibinfo
  {author} {\bibfnamefont {I.~C.}\ \bibnamefont {Mayorga}}, \bibinfo {author}
  {\bibfnamefont {R.}~\bibnamefont {G\"{u}sten}}, \bibinfo {author}
  {\bibfnamefont {J.}~\bibnamefont {Hemberger}},\ and\ \bibinfo {author}
  {\bibfnamefont {M.}~\bibnamefont {Gr\"{u}ninger}},\ }\bibfield  {title}
  {\enquote {\bibinfo {title} {Using a fiber stretcher as a fast phase
  modulator in a continuous wave terahertz spectrometer},}\ }\href
  {https://doi.org/10.1364/JOSAB.29.000614} {\bibfield  {journal} {\bibinfo
  {journal} {J. Opt. Soc. Am. B}\ }\textbf {\bibinfo {volume} {29}},\ \bibinfo
  {pages} {614--620} (\bibinfo {year} {2012})}\BibitemShut {NoStop}%
\bibitem [{TOP()}]{TOPAS}%
  \BibitemOpen
  \href@noop {} {\enquote {\bibinfo {title}
  {https://topas.com/products/topas-coc-polymers/}}\ }\BibitemShut {NoStop}%
\bibitem [{\citenamefont {Mičica}\ \emph {et~al.}(2016)\citenamefont
  {Mičica}, \citenamefont {Bucko}, \citenamefont {Postava}, \citenamefont
  {Vanwolleghem}, \citenamefont {Lampin},\ and\ \citenamefont
  {Pištora}}]{Micica2016}%
  \BibitemOpen
  \bibfield  {author} {\bibinfo {author} {\bibfnamefont {M.}~\bibnamefont
  {Mičica}}, \bibinfo {author} {\bibfnamefont {V.}~\bibnamefont {Bucko}},
  \bibinfo {author} {\bibfnamefont {K.}~\bibnamefont {Postava}}, \bibinfo
  {author} {\bibfnamefont {M.}~\bibnamefont {Vanwolleghem}}, \bibinfo {author}
  {\bibfnamefont {J.-F.}\ \bibnamefont {Lampin}},\ and\ \bibinfo {author}
  {\bibfnamefont {J.}~\bibnamefont {Pištora}},\ }\bibfield  {title} {\enquote
  {\bibinfo {title} {Analysis of wire-grid polarisers in terahertz spectral
  range},}\ }\href {https://doi.org/10.1166/jnn.2016.12556} {\bibfield
  {journal} {\bibinfo  {journal} {Journal of Nanoscience and Nanotechnology}\
  }\textbf {\bibinfo {volume} {16}},\ \bibinfo {pages} {7810--7813} (\bibinfo
  {year} {2016})}\BibitemShut {NoStop}%
\bibitem [{\citenamefont
  {https://www.plxinc.com/products/hollow-roof-mirrors-hrm/groups/rm
  202}()}]{Rooftop}%
  \BibitemOpen
  \bibfield  {author} {\bibinfo {author} {\bibnamefont
  {https://www.plxinc.com/products/hollow-roof-mirrors-hrm/groups/rm 202}}\
  }\href@noop {} {}\BibitemShut {NoStop}%
\bibitem [{\citenamefont
  {https://www.thorlabs.com/thorproduct.cfm?partnumber=MT1/M
  Z8}()}]{Thorlabs12mm}%
  \BibitemOpen
  \bibfield  {author} {\bibinfo {author} {\bibnamefont
  {https://www.thorlabs.com/thorproduct.cfm?partnumber=MT1/M Z8}}\ }\href@noop
  {} {}\BibitemShut {NoStop}%
\bibitem [{\citenamefont {Feil}\ and\ \citenamefont {Haas}(1987)}]{Feil1987}%
  \BibitemOpen
  \bibfield  {author} {\bibinfo {author} {\bibfnamefont {H.}~\bibnamefont
  {Feil}}\ and\ \bibinfo {author} {\bibfnamefont {C.}~\bibnamefont {Haas}},\
  }\bibfield  {title} {\enquote {\bibinfo {title} {Magneto-optical {Kerr}
  effect, enhanced by the plasma resonance of charge carriers},}\ }\href
  {https://doi.org/10.1103/PhysRevLett.58.65} {\bibfield  {journal} {\bibinfo
  {journal} {Phys. Rev. Lett.}\ }\textbf {\bibinfo {volume} {58}},\ \bibinfo
  {pages} {65--68} (\bibinfo {year} {1987})}\BibitemShut {NoStop}%
\bibitem [{\citenamefont {Howells}\ and\ \citenamefont
  {Schlie}(1996)}]{Howells1996}%
  \BibitemOpen
  \bibfield  {author} {\bibinfo {author} {\bibfnamefont {S.~C.}\ \bibnamefont
  {Howells}}\ and\ \bibinfo {author} {\bibfnamefont {L.~A.}\ \bibnamefont
  {Schlie}},\ }\bibfield  {title} {\enquote {\bibinfo {title} {Transient
  terahertz reflection spectroscopy of undoped {InSb} from 0.1 to 1.1 {THz}},}\
  }\href {https://doi.org/10.1063/1.117783} {\bibfield  {journal} {\bibinfo
  {journal} {Applied Physics Letters}\ }\textbf {\bibinfo {volume} {69}},\
  \bibinfo {pages} {550--552} (\bibinfo {year} {1996})}\BibitemShut {NoStop}%
\end{thebibliography}
%

\end{document}